\documentclass[twocolumn]{aastex631}

\usepackage{stix}
\usepackage{amsmath}
\usepackage{multirow}
\usepackage{savesym}
\usepackage{upgreek}
\savesymbol{tablenum}
\usepackage{siunitx}
\restoresymbol{SIX}{tablenum}
\newcommand{\as}[1]{\ang[angle-symbol-over-decimal]{;;#1}}
\setcitestyle{notesep={ }}

\usepackage{colortbl}

\shorttitle{Callisto Properties from ALMA Calibration Data}
\shortauthors{Meyer et al.}

\begin{document}

\title{Revealing Callisto’s Near Subsurface Thermophysical Properties with ALMA Calibration Data}

\correspondingauthor{Cole Meyer}
\email{cmmeyer@arizona.edu}
\author[0009-0006-2781-3484]{Cole Meyer}
\affiliation{Lunar and Planetary Laboratory, University of Arizona, Tucson, AZ 85721, USA}
\affiliation{Department of Astrophysical Sciences, Princeton University, Princeton, NJ 08544, USA}

\author[0000-0003-3887-4080]{Maria Camarca}
\affiliation{Division of Geological and Planetary Sciences, California Institute of Technology, Pasadena, CA 91125, USA}

\author[0000-0002-9068-3428]{Katherine de Kleer}
\affiliation{Division of Geological and Planetary Sciences, California Institute of Technology, Pasadena, CA 91125, USA}

\author[0000-0002-8178-1042]{Alexander Thelen}
\affiliation{Division of Geological and Planetary Sciences, California Institute of Technology, Pasadena, CA 91125, USA}

\author[0000-0002-6757-4522]{Christopher Chyba}
\affiliation{Department of Astrophysical Sciences, Princeton University, Princeton, NJ 08544, USA}
\affiliation{School of Public and International Affairs, Princeton University, Princeton, NJ 08544, USA}

\author[0000-0002-5344-820X]{Bryan Butler}
\affiliation{National Radio Astronomy Observatory, Socorro, NM 87801, USA}




\begin{abstract}

Thermal images at different wavelengths probe varying subsurface depths of planetary bodies, and therefore can inform us about their compositions, thermophysical properties, and impact histories. We identified six archival observations of Callisto obtained by the Atacama Large Millimeter/submillimeter Array (ALMA) between 2012 July 17 and 2012 November 4 at wavelengths of 0.43--0.47 mm (701.9--641.5 GHz). These wavelengths are shorter than those of nearly all other Callisto ALMA data and are sensitive to subsurface emission at depths (the upper $\sim$cm) between those sounded by millimeter and infrared observations. We estimate the disk-averaged brightness temperature as $133\pm15$ K, and use a thermophysical mixture model to find that Callisto's thermal emission is best fit by a $\sim$50--50\% two-component thermal inertia mixture of $\Gamma_\text{low}$$\sim$15--40 and $\Gamma_\text{high}$$\sim$1200--2000 J m$^{-2}$ K$^{-1}$ s$^{-1/2}$, consistent with recent ALMA observations of Callisto at longer wavelengths. Finally, we present several previously unpublished \textit{Galileo} Photopolarimeter-Radiometer (PPR) observations of Callisto and derive thermal inertia and spectral emissivity maps using the same model. Altogether, these ALMA and PPR maps improve our understanding of the thermal properties and spatial distribution of Callisto’s shallow subsurface regolith, and demonstrate the value of ALMA flux density calibrator data for extending frequency coverage of existing science data.

\end{abstract}

\keywords{Callisto (2279); Galilean satellites (627); Jovian satellites (872); Radio interferometry (1346); Planetary surfaces (2113); Surface processes (2116); Planetary science (1255); Natural satellite surfaces (2208); Remote sensing (2191); Millimeter astronomy (1061); Submillimeter astronomy (1647)}


\section{Introduction} \label{section:intro}

The archetype of geologic quiescence, Callisto retains one of the oldest surfaces~\citep[$\sim$4.5 Gyr;][]{moore_callisto_2004} in the solar system and offers an excellent record of long-term impact modification. The outermost of Jupiter's Galilean satellites, Callisto hosts a heavily cratered surface which contrasts sharp-crested bright, frosty rims against dark, smooth plains and low-lying regions far from geologic relief~\citep{spencer_surfaces_1987}. These disparate terrains link the largest multi-ring impact basins in the solar system, which occupy much of Callisto's surface area and place key constraints on the properties of its old surface. Although a subsurface ocean may surround its partially differentiated interior~\citep{zimmer_subsurface_2000, anderson_distribution_1998}, Callisto bears little to no evidence of endogenic geologic activity due in part to its exclusion from the Laplace resonance that tidally heats its three Galilean siblings. Therefore, Callisto provides a baseline against which to compare the tidally-driven evolutionary pathways of Io, Europa, and Ganymede~\citep{moore_callisto_2004, schenk_geology_1995}.

Callisto's bulk density is the lowest of the Galileans~\citep[$\sim$1.8 g cm$^{-3}$, with $\sim$1 g cm$^{-3}$ and $\sim$3 g cm$^{-3}$ corresponding to solid water ice and rock, respectively;][]{morrison_satellites_1982}, a consequence of its high water ice content~\citep[e.g.,][]{pilcher_galilean_1972, kieffer_frost_1974}. At the same time, its low albedo (0.2) and subsequent high surface temperatures indicate the presence of dark, non-ice surface materials~\citep{moore_callisto_2004,greeley_galileo_2000}. The composition and origin of these dark materials remain unclear, with endogenic theories ranging from rim-forming bedrock decomposition to volatile sublimation~\citep{moore_mass_1999,chuang_large_2000} and exogenic theories typically invoking meteoroid or dust infall from the irregular Jovian satellites~\citep{bottke_black_2013,chen_life_2024}. Weakening of crater walls by sublimation erosion of the volatile-rich subsurface material coupled with seismic triggering by nearby impacts appears to liberate dark, rim-forming material (e.g., bedrock) that subsequently fills crater floors~\citep{moore_callisto_2004}. That these dark materials have higher abundances (and thus, that large-scale mass wasting occurs more frequently) on Callisto than its Galilean siblings may suggest fundamental differences in crustal properties and surface destabilization processes which are not well understood~\citep{moore_mass_1999,chuang_large_2000}. Altogether, these features underscore our limited understanding of Callisto's inactive surface processes and motivate further analysis of its characteristics.

As airless bodies like the icy Galileans endure diurnal cycles and eclipses, a thermal wave driven by variable solar insolation propagates through their uppermost layers~\citep{ferrari_thermal_2018}. Consequently, their surfaces radiate heat into space as variable thermal emission at infrared and (sub)millimeter wavelengths, which can be used alongside heat transfer models to infer thermophysical properties such as spectral emissivity ($\epsilon$) and thermal inertia~\citep[$\Gamma$;][]{ferrari_thermal_2018,spencer_surfaces_1987}. These inferred thermal properties are in turn informed by physical characteristics of the surface regolith such as ice phase, grain size, roughness and porosity, and the vertical compaction profile~\citep[i.e., density/porosity as a function of depth;][]{de_kleer_ganymedes_2021,ferrari_thermal_2018}. Therefore, the thermal emission emanating from the (sub)surfaces of airless, icy bodies hints at their past geologic evolution, surface regolith composition and spatial distribution, and internal structure. For example,~\citet{spencer_temperatures_1999} discovered an equatorial night-time temperature anomaly on Europa uncorrelated with surface albedo or geology using the Photopolarimeter-Radiometer (PPR) aboard \textit{Galileo}. Later, \citet{rathbun_galileo_2010} used the vertically homogeneous thermal model of~\citet{spencer_systematic_1989} to show that the anomaly could be explained by latitudinal variation in thermal inertia. Similarly, infrared spectra from the \textit{Galileo} Near-Infrared Spectrometer (NIMS) indicated a dominance of crystalline and amorphous water ice on the immediate surface of Callisto and Europa, respectively, with latitudinal variation in phase across Ganymede's surface~\citep{hansen_amorphous_2004}. Such a distribution could be explained by radiolytic disruption at small orbital radii, whereas Callisto may experience increased thermal kinetic crystallization due to its relatively high surface ice temperatures~\citep{hansen_amorphous_2004,stephan_h2o-ice_2020}.

On the ground, the Atacama Large Millimeter/submillimeter Array (ALMA) offers a complementary perspective with high sensitivity and resolution at slightly longer wavelengths~\citep[$\sim$0.4--8.6 mm;][]{cortes_alma_2024} than PPR and NIMS~\citep[$\sim$15--100 and $\sim$0.7--5.2 $\upmu$m, respectively;][]{carlson_near-infrared_1992,russell_galileo_1992}. The depth at which observed thermal emission effectively originates depends in part upon the wavelength of observation, where infrared and (sub)millimeter wavelengths probe the upper tens of microns to meters on icy satellite surfaces, respectively~\citep{ferrari_thermal_2018,de_kleer_ganymedes_2021}. Therefore, multi-wavelength thermal observations allow the construction of an effective temperature depth profile, yielding additional insights into the balance between different endo- and exogenic processes shaping the structure of the subsurface. For example,~\citet{trumbo_alma_2017} applied a global thermal model to an ALMA daytime image of Europa near Pwyll crater, the aforementioned region of \textit{Galileo} PPR nighttime thermal excess~\citep{spencer_temperatures_1999} later associated with two potential plume detections~\citep{sparks_probing_2016,sparks_active_2017}, to show that a moderate increase in local thermal inertia was sufficient to produce the excess. Soon after,~\citet{trumbo_alma_2018} supplemented the first ALMA image with several more to obtain surface-wide thermal inertia and spectral emissivity maps at 233 GHz (1.3 mm), and identify an anomalously cool region near 90$^\circ$W and 23$^\circ$N. Recently,~\citet{thelen_subsurface_2024} applied a more sophisticated thermophysical model (see Section~\ref{section:model} for details) to ALMA images of Europa across several frequency bands (97.5, 233, and 343.5 GHz, corresponding to 3.05, 1.25, and 0.88 mm, respectively) to obtain porosities and their associated effective thermal inertiae at three different near-subsurface depths. Additionally, the images exhibited thermal anomalies co-located with several geologic surface features including the expansive rays of Pwyll crater observed by the aforementioned authors. Previously, \citet{de_kleer_ganymedes_2021} found \textit{Galileo} PPR infrared data in combination with ALMA (sub)millimeter data of Ganymede to be consistent with a compaction profile for which porosity drops from $\sim$85\% at the surface to 10$^{+30}_{-10}$\% over a compaction length scale of tens of centimeters. Altogether, these results were made possible by the variation in inferred thermal properties with wavelength and therefore with effective depth of origin.

Recent work by~\citet{camarca_thermal_2023} and~\citet{camarca_multifrequency_2025} extends these (sub)millimeter analyses to Callisto at ALMA wavelengths of 0.87 mm (343 GHz) to 3 mm (97 GHz) using the thermophysical model of~\citet{de_kleer_ganymedes_2021} and~\citet{thelen_subsurface_2024}. They achieve optimal fits using a model with a linear mixture of terrains with two different thermal inertiae (henceforth referred to as a ``two-$\Gamma$'' model) and identify thermal anomalies across the surface consistent with micrometeorite bombardment on the leading hemisphere and possible variation in ice composition on the trailing. These wavelengths are sensitive to emission from the upper $\sim$5--100 cm of surface, whereas shorter ALMA and infrared wavelengths sense the upper tens of $\upmu$m to cm of the surface. Additional observations using ALMA or other facilities at sub-mm wavelengths could constrain near-surface ($\sim$1--5 cm) thermal properties of Valhalla or different geologic terrains and complete depth coverage bookended by existing ALMA data and infrared space-based data.

We present six ALMA observations of Callisto at 0.47 mm (641.5 GHz) to 0.43 mm (701.9 GHz) that span all sub-observer longitudes and sense the upper $\sim$10 mm of the surface. These data, their corresponding observational parameters, and the use of archival calibrator data are described in Section~\ref{section:methods}. In Section~\ref{section:model}, we outline the thermophysical model~\citep{de_kleer_ganymedes_2021} and its application to fitting the surface with linear mixtures of different surface properties~\citep{camarca_thermal_2023}. The resulting residual maps and inferred thermal properties are presented in Section~\ref{section:results} alongside a brief presentation and analysis of several unpublished \textit{Galileo} PPR observations of Callisto at 16.8 $\upmu$m and 27.5 $\upmu$m. Finally, our conclusions are presented in Section~\ref{section:conclusion}.

\section{Methods}\label{section:methods}
\subsection{Observations}\label{subsection:observations}
The relative geologic inactivity of Callisto makes it an effective flux density calibration object, and there exist public calibrator data of Callisto in the ALMA Science Archive\footnote{\href{https://almascience.nrao.edu/aq/}{https://almascience.nrao.edu/aq/}} across nearly all receiver bands, including those in which no Callisto science data exist. We selected flux density calibrator observations with spatial resolutions sufficient to resolve its major surface features ($\sim$750 km) and identified six observations obtained between 2012 July 16 and 2012 November 4 that together have full longitudinal coverage of Callisto's surface. These data are associated with Programs 2011.0.00199.S (PI Tomoya Hirota), 2011.0.00647.S (PI Claudio Codella), and 2011.0.00223.S (PI Valentin Bujarrabal). The observations were taken using the Band 9 receivers at central frequencies between 641.5 and 701.9 GHz (wavelengths of 0.46 and 0.43 mm, respectively), each with four spectral windows collectively spanning 937.5 MHz of total bandwidth. Since these are ALMA Cycle 0 observations, only 21 to 28 12 m antennas were used with baselines ranging from 21 to 402 m and on-source integration times between 191 and 321 s. Other details of the observations, including angular and spatial resolutions, viewing geometries, disk-integrated flux densities, and disk-averaged brightness temperatures (see Section~\ref{section:fluxcal}) are detailed in Table~\ref{table:obsparams}.

\begin{figure*}[t]
\centering
\includegraphics[width=0.78\textwidth]{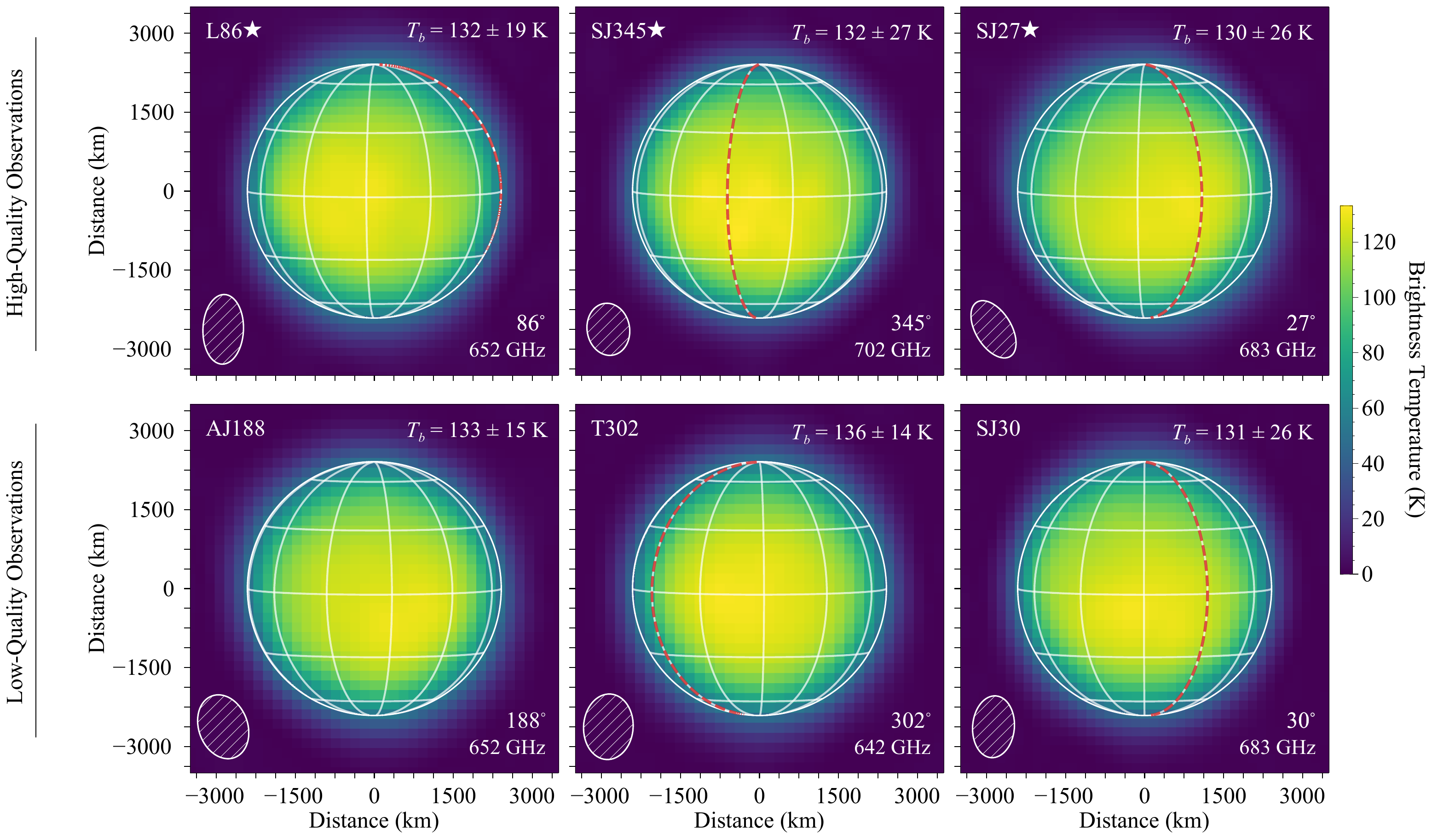}

\caption{ALMA brightness temperature maps of Callisto separated into ``high-quality'' (top row) and ``low-quality'' observations to encourage caution during interpretation. The observation code, sub-observer longitude, and observation frequency are labeled, and the 2D full-width at half maximum of the ALMA point spread function (representing the effective resolution element) is shown in the bottom left of each subplot. The central prime meridian ($l=0^\circ$) is shown as a red dashed line and white lines indicate latitudes and longitudes with 30$^\circ$ spacing. All images use the same color scale and are oriented such that Callisto's north pole aligns with the vertical axis.}
\label{fig:init_ims}
\end{figure*}

Observations from ALMA Cycle 0 were obtained during ongoing commissioning of the observatory and thus must be carefully reduced. Array configurations from ALMA early science employed fewer antennas with shorter baselines than those of recent cycles, leading to diminished $uv$ coverage, or equivalently, sparser sampling of the Fourier transform of the sky brightness distribution. As such, deconvolved images may exhibit increased noise, image artifacts, and incomplete sampling of certain spatial scales. Moreover, quasar calibrators tend to be weak in Band 9 and were infrequently observed during Cycle 0 operations. Techniques to mitigate the resulting bandpass, phase, and amplitude calibration deficiencies had not yet been developed. Fortunately, our observations exhibit relatively few such deficiencies compared to other Cycle 0 observations. That said, three observations presented here contain potentially problematic characteristics and should be cautiously interpreted. In particular, AJ188 (see Table~\ref{table:obsparams} for code definition) likely contains sidelobe contamination from Jupiter, which is separated from Callisto by $\sim$45$^{\prime\prime}$ at the time of observation (observation parameters defined in Table~\ref{table:obsparams}). The half-width between nulls of the antenna primary beam in Band 9 is small ($\sim$\as{8.5}), however peaks exist outside the central beam within the AJ188 point spread function, resulting in increased noise and possible artifacts in the deconvolved image. Moreover, bandpass calibration for T302 and SJ30 was poor and multiple antennas were flagged due to oscillating system temperatures (internal communication, ALMA staff). The resulting $uv$ coverage suffered, and likely resulted in the observed bright artifacts along their top edges (see Figure~\ref{fig:init_ims}). Atmospheric absorption features and faulty antennas were flagged in the remaining observations, and the corresponding deconvolved images represent Callisto's true emission. We append a star symbol ($\bigstar$) to the codes that correspond to these three ``high-quality observations'' (see Table~\ref{table:obsparams}) and present the two groups of deconvolved images separately (see Figure~\ref{fig:init_ims}) to encourage caution during interpretation. Note that the terms ``high'' and ``low'' quality are not conditions on an absolute scale, but rather relative between the presented observations.

\begin{deluxetable*}{cccccccccccc}
\tablenum{1}
\tabletypesize{\scriptsize}
\tablecaption{Observational Parameters and Derived Properties\label{table:obsparams}}
\tablewidth{0pt}
\tablehead{
\colhead{Code$^a$} & \colhead{Date} &
\colhead{Time} & \colhead{Ang. diam.} &
\colhead{Beam} & \colhead{Sub-obs. lat.} &
\colhead{Sub-obs. lon.} & \colhead{Phase ang.} &
\colhead{$\nu^b$} & \colhead{$\lambda$} &
\colhead{$F_\nu$} & \colhead{$T_b$} \\[-5pt]
\colhead{} & \colhead{(UT)} &
\colhead{(UT)} & \colhead{($\prime\prime$)} &
\colhead{($\prime\prime$)} & \colhead{($^\circ$N)} &
\colhead{($^\circ$W)} & \colhead{($^\circ$)} &
\colhead{(GHz)} & \colhead{(mm)} &
\colhead{(Jy)} & \colhead{(K)}
}
\startdata
AJ188 & 2012 Jul 17 & 12:48 & 1.18 & $0.31\times0.23$ & 2.8 & 188.0 & 8.8 & 651.5 & 0.46 & $39\pm5$ & $133\pm15$ \\
T302 & 2012 Aug 25 & 11:33 & 1.30 & $0.34\times0.25$ & 2.8 & 302.1 & 11.4 & 641.5 & 0.47 & $48\pm6$ & $136\pm14$ \\
L86$\bigstar$ & 2012 Oct 21 & 08:13 & 1.54 & $0.42\times0.25$ & 2.8 & 86.4 & 8.4 & 651.5 & 0.46 & $67\pm11$ & $132\pm19$ \\
SJ345$\bigstar$ & 2012 Nov 02 & 07:04 & 1.58 & $0.33\times0.26$ & 2.8 & 345.3 & 6.4 & 701.9 & 0.43 & $80\pm19$ & $132\pm27$ \\
SJ27$\bigstar$ & 2012 Nov 04 & 05:09 & 1.59 & $0.40\times0.22$ & 2.8 & 26.8 & 6.1 & 682.8 & 0.44 & $76\pm17$ & $130\pm26$ \\
SJ30 & 2012 Nov 04 & 07:58 & 1.59 & $0.39\times0.26$ & 2.8 & 29.7 & 6.1 & 682.8 & 0.44 & $76\pm18$ & $131\pm26$ \\
\enddata
\tablenotetext{a}{Code constructed by joining hemisphere abbreviation (L$=$leading, T$=$trailing, SJ$=$subjovian, AJ$=$antijovian) and sub-observer longitude ($^\circ$W). The star symbol ($\bigstar$) indicates a high-quality observation, as described in Section~\ref{subsection:observations}.}
\tablenotetext{b}{Central frequency determined by averaging frequencies of each spectral window. All frequencies correspond to ALMA Band 9 (602 to 720 GHz).}
\end{deluxetable*}

\begin{deluxetable*}{ccccccccccc}
\tablenum{2}
\tablecaption{Flux Density Calibration Results Using \texttt{getALMAFlux} and Statistical Estimation (SE)\label{table:fluxcal}}
\tablewidth{0pt}
\tablehead{
\multicolumn{4}{c|}{Callisto Observations} & \multicolumn{7}{c}{Quasar Observations} \\[10pt]
\cline{5-11}
\multicolumn{4}{c|}{} & \multicolumn{3}{c|}{\texttt{getALMAFlux} Spectral Index} & \multicolumn{2}{c|}{SE Spectral Index} & \multicolumn{2}{c}{Extrapolation} \\[10pt]
\cline{1-11}
\colhead{Code$^a$} & \colhead{Obs. date} & \colhead{$\nu^b$} & \multicolumn{1}{c|}{Quasar} & \colhead{Obs. date} & \colhead{$\nu^c$} & \multicolumn{1}{c|}{$\alpha$} & \colhead{Num. Obs.} & \multicolumn{1}{c|}{14-day $\alpha$} & \colhead{\texttt{getALMAFlux} $F_\nu$} & \colhead{SE $F_\nu$}
\\[-5pt]
\colhead{} & \colhead{(UT)}  & \colhead{(GHz)} & \multicolumn{1}{c|}{} & \colhead{(UT)} & \colhead{(GHz)} & \multicolumn{1}{c|}{} & \colhead{} & \multicolumn{1}{c|}{} & \colhead{(Jy)} & \colhead{(Jy)}
}
\startdata
AJ188 & 2012 Jul 17 & 651.5 & J0423--0120 & 2012 Jun 30 & 221.0$^6$ & --1.13 $\pm$ 0.63 & 2019 & --0.56 $\pm$ 0.09  & 1.15 $\pm$ 0.55 & 1.66 $\pm$ 0.21\\
&&&& 2012 Jun 30 & 343.3$^7$ &&&&& \\
\\[-5pt]
T302 & 2012 Aug 25 & 641.5 & J0423--0120 & 2012 Aug 26 & 221.0$^6$ & --0.75 $\pm$ 0.01 & 2019 & --0.74 $\pm$ 0.09  & 2.12 $\pm$ 0.14 & 2.13 $\pm$ 0.25 \\
&&&& 2012 Aug 24 & 343.3$^7$ &&&&& \\
\\[-5pt]
L86$\bigstar$ & 2012 Oct 21 & 651.5 & J0423--0120 & \textbf{2012 Aug 26} & \textbf{221.0$^6$} & \textbf{--0.75 $\pm$ 0.01} & 2019 & --0.62 $\pm$ 0.09 & \textbf{1.60 $\pm$ 0.12} & 1.84 $\pm$ 0.29 \\
&&&& \textbf{2012 Aug 24} & \textbf{343.3$^7$} &  &&&  & \\
\\[-5pt]
SJ345$\bigstar$ & 2012 Nov 02 & 701.9 & J0538-4405 & 2012 Oct 06 & 221.0$^6$ & --0.74 $\pm$ 0.12 & 1897 & --0.70 $\pm$ 0.11 & 0.87 $\pm$ 0.22 & 0.94 $\pm$ 0.22 \\
&&&& 2012 Oct 06 & 343.3$^7$ &&&&& \\
\\[-5pt]
SJ27$\bigstar$ & 2012 Nov 04 & 682.8 & J0319+4130 & \textbf{2013 Jun 17} & \textbf{98.2$^3$} & \textbf{--0.79 $\pm$ 0.05} & 693 & --0.70 $\pm$ 0.11 & \textbf{3.84 $\pm$ 0.53} & 0.96 $\pm$ 0.22 \\
&&&& \textbf{2013 Jun 17} & \textbf{349.3$^7$} &  &&&  & \\[5pt]
SJ30 & 2012 Nov 04 & 682.8 & J0538-4405 & 2012 Oct 06 & 221.0$^6$ & --0.74 $\pm$ 0.12 & 1897 & --0.70 $\pm$ 0.11 & 0.89 $\pm$ 0.22 & 0.96 $\pm$ 0.22 \\
&&&& 2012 Oct 06 & 343.3$^7$ &&&&& \\
\\
\enddata
\tablenotetext{a}{Code constructed by joining hemisphere abbreviation (L$=$leading, T$=$trailing, SJ$=$subjovian, AJ$=$antijovian) and sub-observer longitude ($^\circ$W). The star symbol ($\bigstar$) indicates a high-quality observation, as described in Section~\ref{subsection:observations}.}
\tablenotetext{b}{Central frequency determined by averaging frequencies of each spectral window. All frequencies correspond to ALMA Band 9 (602 to 720 GHz).}
\tablenotetext{c}{Each exponent indicates which ALMA receiver band the corresponding frequency falls under (Band 3: 84 to 116 GHz; Band 6: 211 to 275 GHz; Band 7: 275 to 373 GHz).}
\tablecomments{Bold font indicates quasar measurements that occurred at least 30 days from the Callisto observation date.}
\end{deluxetable*}

\subsection{Data reduction}

The raw data obtained by ALMA were reduced and calibrated using the Common Astronomy Software Applications (CASA) package ver. 5.6.1~\citep{the_casa_team_casa_2022}, resulting in a calibrated set of complex ``visibilities'' (see Section~\ref{section:fluxcal} for flux density calibration). These visibilities are fundamentally the Fourier transform of the sky brightness distribution sampled at discrete spatial frequencies, and can be used to reconstruct both spatial and spectral structure using image deconvolution techniques. For an in-depth review of interferometric imaging of solar system objects (SSOs), see~\citet{butler_solar_1999}. The calibrated visibilities were flagged for telluric contamination and averaged into 256 MHz spectral bins to reduce data volume. Following recent ALMA SSO imaging~\citep[e.g.,][]{thelen_subsurface_2024,camarca_multifrequency_2025,de_kleer_ganymedes_2021}, we employ multi-frequency synthesis~\citep{conway_multi-frequency_1990} and iterative self-calibration~\citep{cornwell_self-calibration_1999} imaging procedures to improve phase coherence and signal-to-noise ratios (S/N). We use a uniform limb-darkened disk as the \texttt{startmodel} for phase-only self-calibration to avoid amplifying noise spikes, as is standard for imaging high S/N planetary objects. See~\citet{brogan_advanced_2018} for a detailed outline of the self-calibration procedure.

The visibility function is only partially sampled due to incomplete antenna ($uv$) coverage, and image deconvolution techniques must be employed to remove interferometric artifacts and effectively reconstruct the sky brightness distribution. We employ the CASA \texttt{tclean}~\citep{rau_multi-scale_2011} task for deconvolution, using a Briggs weighting scheme with a ``robust'' parameter of 0.5, gain of 0.05, and the Högbom deconvolver~\citep{hogbom_aperture_1974}. As a start model, we used a limb-darkened disk the size and brightness of Callisto. A flux threshold of twice the expected RMS (root mean square) noise (typically on order 20 mJy) was used, and a primary beam correction was applied to the resulting image products. The RMS noise was measured from a non source region of the non primary beam-corrected image product. The disk-integrated flux density for each deconvolved image and final calibrated images are presented in Table~\ref{table:obsparams} and Figure~\ref{fig:init_ims}, respectively. See Section~\ref{section:dabt} for a discussion of our procedure for obtaining disk-integrated flux densities.

\subsection{Flux density calibration}\label{section:fluxcal}

The measured visibility amplitudes reported by ALMA are given as the ratio between the measured signal and the total system noise~\citep{cortes_alma_2024}. To convert the signal to physically meaningful units of Janskys (Jy), ALMA observes a bright, stable calibrator object whose flux density can be accurately predicted using a model. By comparing the visibility amplitudes to the flux density predicted by the model, a scaling factor is obtained and applied to the data, thereby setting the flux density scale. In the case of our Callisto calibrator observations (prior to the re-calibration described below), the flux density of Callisto (in arbitrary units) was compared with its corresponding ``Butler-JPL-Horizons 2012''~\citep{butler_alma_2012} model to obtain a scaling factor, which was subsequently applied to the other calibrator and science targets. In other words, the prior flux density of Callisto was initially set by a model and cannot be used for analysis as is. To glean any information about its true flux density, its flux density scale must be reset independently using a different object.

Quasars are routinely observed by ALMA for the purpose of array pointing, phase, and bandpass response calibration. Fortunately, especially bright and relatively stable quasars are also used as flux density calibrators when solar system calibrators are not visible in the sky. To measure their intrinsic brightness variability, near-in-time measurements of their flux density are interpolated to the date and frequency of the calibrator observation. The flux density scale is set by comparing the interpolated flux density with the measured flux density. In recent ALMA cycles, flux densities of the brightest, uniformly distributed ``grid source'' quasars were obtained at least every two weeks across receiver Bands 3 (84--116 GHz), 6 (211--275 GHz), and 7~\citep[275--373 GHz; see Chapter 10 of][for more details]{cortes_alma_2024}. However, it is clear from a review of archival ALMA calibrator observations that early ALMA operations did not see the same regularity in flux density measurements, so care must be taken to interpolate only across timescales on which quasar brightness is relatively stable~\citep[$\sim$two weeks;][]{guzman_stochastic_2019}.

Given the relative irregularity of flux density measurements during Cycle 0, we estimate the true flux density of our new quasar flux density calibrators using two different methods. First, the \texttt{getALMAFlux} task from the \texttt{analysisUtils} Python package searches the ALMA Calibrator Source Catalogue\footnote{\href{https://almascience.eso.org/alma-data/calibrator-catalogue}{https://almascience.eso.org/alma-data/calibrator-catalogue}} for two near-in-time observations in different receiver bands. These observations are fit to a quasar spectral profile of the form
\begin{equation}
    \left(\frac{F_{\nu,1}}{F_{\nu,2}}\right) = \left(\frac{\nu_1}{\nu_2}\right)^\alpha,
\end{equation}
where $F_\nu$ is the flux density, $\nu$ is the observation frequency, and $\alpha$ is the spectral index~\citep[typically between \mbox{--0.35} and \mbox{--0.80} for ALMA calibrators;][]{guzman_stochastic_2019}. Finally, the derived spectral index is used to extrapolate the flux density to the observation frequency, and a Monte-Carlo simulation used to estimate the uncertainty. We present the quasar observation dates and frequencies, \texttt{getALMAFlux} spectral index, and extrapolated flux density for each Callisto observation in Table~\ref{table:fluxcal}. Note that the ALMA pipeline now employs the Flux Estimation Service~\citep{kneissl_flux_2022}, which uses the Levenberg-Marquardt algorithm~\citep{marquardt_algorithm_1963,levenberg_method_1944} to improve the overall fit to the spectral profile. However, too few quasar observations occurred near our observation dates for the algorithm to converge.

\begin{figure}
\centering
\includegraphics[width=0.45\textwidth]{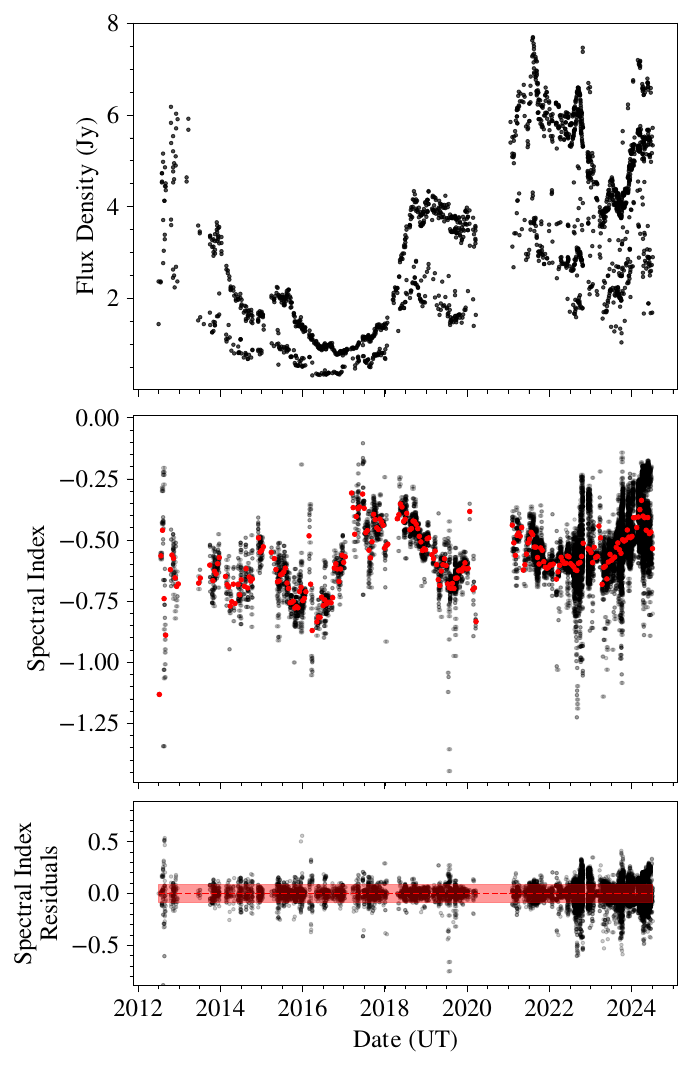}
\caption{Example of quasar spectral index estimation for J0423-0120. Top panel: flux density over time. Middle panel: in black points are the individual spectral indices, with the 14-day $\alpha$ shown in red. Bottom panel: black points are the 14-day $\alpha$ residuals, and the red envelope identifies one standard deviation in the residuals.}
\label{fig:quasar}
\end{figure}

The challenge with using \texttt{getALMAFlux} for Cycle 0 observations is its reliance on near-in-time observations to estimate the spectral index. Ideally, measurements in Bands 3 and 7 would be used for combined small weather effects, high sensitivity, and a large frequency lever arm~\citep{guzman_stochastic_2019}. However, quasars were scarcely observed in Band 3 near our observation dates so \texttt{getALMAFlux} was forced to use measurements in Bands 6 and 7, thereby increasing the error on the spectral index. To remedy this, we also explored statistical estimation (SE) of the spectral index and its uncertainty. Specifically, we obtained flux density measurements for the relevant quasars from the ALMA Calibrator Source Catalogue across all available frequencies and dates. For each measurement, we calculated spectral indices between it and all other measurements within 14 days (the timescale on which quasar brightness is typically stable) on either side of the observation date. Note that we exclude spectral indices outside the --1.5 to 0 range in accordance with stochastic modeling of these quasars~\citep[e.g.,][]{guzman_stochastic_2019}. Next, we separated the spectral indices into adjacent 14-day bins and averaged each bin to obtain an estimate for the spectral index every 14 days (henceforth referred to as the 14-day $\alpha$), where the 14-day $\alpha$ for a given observation is the one nearest in time to the observation date. To determine the uncertainty on the 14-day $\alpha$ estimates, we subtracted each 14-day $\alpha$ from the spectral indices in its corresponding 14-day bin to obtain a set of residuals. The standard deviation of the 14-day $\alpha$ residuals represents the ability for the 14-day $\alpha$ to recover the measured flux densities, and therefore is the uncertainty on these $\alpha$ estimates. The number of quasar observations used for uncertainty estimation, 14-day $\alpha$, and extrapolated flux density for each Callisto observation are presented in Table~\ref{table:fluxcal}. An example of quasar spectral index estimation for J0423-0120 is given by Figure~\ref{fig:quasar}.

The only instance in which \texttt{getALMAFlux} retrieved quasar observations within 14 days of the Callisto observation was T302, which indeed exhibited strong agreement in derived properties between the \texttt{getALMAFlux} and SE methods. For the remaining observations, \texttt{getALMAFlux} employed quasar measurements separated from the Callisto observation by up to 225 days, a concerning time lapse given our conservative $\sim$14 day timescale estimate for spectral shape and brightness variability. Accordingly, we employ the more robust SE-derived spectral indices and extrapolated flux densities to set the flux density scale for our Callisto observations, and advise users of \texttt{getALMAFlux} to verify the dates of the quasar measurements it employs~\citep[see][ for an extended discussion of \texttt{getALMAFlux}]{francis_accuracy_2020}.

\subsection{Disk-averaged brightness temperature}~\label{section:dabt}

The ALMA data presented here are intermediate in wavelength between past published ALMA observations and thermal infrared wavelengths. Therefore, to contextualize the relative brightness of Callisto in these observations amongst previous largely unresolved observations of its surface, we derive the disk-integrated flux density $F_\nu$ for each observation. Specifically, we use the CASA \texttt{uvmodelfit} task to fit the complex visibilities with a sombrero function, or the inverse Fourier transform of Callisto's assumed uniform disk. We exclude baselines $>$125 m since longer baselines are noisier and probe smaller-scale spatial structure rather than global emission, the latter of which is needed for $F_\nu$ estimation. The disk-averaged brightness temperature $T_b$ is obtained by assuming that Callisto's thermal emission can be parameterized by the Planck function. In such a case, we invert the following equation to obtain $T_b$:
\begin{equation}
\begin{aligned}
    F_\nu = &10^{26}\frac{\pi R^2_C}{206265^2}\frac{2h\nu^3}{c^2} \\
    &\times \left[\frac{1}{e^{h\nu/k_bT_b}-1}-\frac{1}{e^{h\nu/k_bT_{cmb}}-1}\right],
\end{aligned}
\end{equation}
where $F_\nu$ is in Jy, $R_C$ is the radius of Callisto in arcseconds (corresponding to an assumed physical radius of 2410.3 km), the quantity 206\,265 is the number of arcseconds in a radian, and in SI units, $h$ is the Planck constant, $\nu$ is the observation frequency, $c$ is the speed of light, $k_b$ is the Boltzmann constant, and $T_{cmb}$ is the temperature of cosmic microwave background ($\sim$2.7 K). The quantity 206265 The resulting disk-integrated flux densities and disk-averaged brightness temperatures are reported in Table~\ref{table:obsparams}. Their corresponding uncertainties incorporate spectral index errors (which include quasar flux density measurement error) and small \texttt{uvmodelfit} fitting errors.

\section{Thermophysical Model}\label{section:model}

To simultaneously isolate small-scale brightness temperature variations across Callisto's surface and infer the thermophysical properties of its surface regolith, we employ the thermophysical model described by~\citet{de_kleer_ganymedes_2021} which treats thermal transport by conduction and radiation. Previous analyses of Ganymede~\citep{de_kleer_ganymedes_2021}, Callisto~\citep{camarca_thermal_2023,camarca_multifrequency_2025}, and Europa~\citep{thelen_subsurface_2024} describe the model in detail, and as such, we offer only a brief outline of its functionality. For a grid of latitudes and longitudes atop the satellite, the model evolves an initial exponentially decreasing temperature profile through Callisto's shallow subsurface over time using the 1D heat diffusion equation. The free parameters in this fit are the thermal inertia $\Gamma$ (henceforth in units of J m$^{-2}$ K$^{-1}$ s$^{-1/2}$ for brevity) and spectral emissivity (henceforth referred to as ``emissivity''). We generate models across $\Gamma$ ranging from 15 to 2000, corresponding to very unconsolidated material and water ice, respectively~\citep{ferrari_low_2016}. The bolometric (bond) albedo at each grid point is fixed using measurements from previous spacecraft observations~\citep[see][for more information]{camarca_thermal_2023}, and we adopt similar snow and ice densities, specific heat values, dust-to-ice fractions, and grain sizes to those of Ganymede and Europa, where appropriate. For a robust exploration of the influence of these fixed parameters on the resulting models, see Section 3 of~\citet{de_kleer_ganymedes_2021}.

The initial temperature profile was evolved over 1000 time steps per Callisto day (1 Callisto day $=$ 16.69 Earth days) for up to $\sim$15 days before convergence, and extended several diurnal thermal skin depths (the depth across which the thermal wave is attenuated by a factor of 1/$e$) beneath the surface. This approach accounts for emission from the subsurface, which ALMA is sensitive to. Since cold, icy surfaces like that enclosing Callisto are highly transparent to (sub)millimeter emission, we integrated the resulting profile along the line of sight in accordance with the dielectric properties to produce a disk model not yet convolved with the ALMA synthesized beam. Following~\citet{camarca_thermal_2023}, we generated mixture models for each combination of $\Gamma_{\text{low}}$ (15 to 400) and $\Gamma_{\text{high}}$ (500 to 2000) model using a weighted sum. The result was a set of 11 mixture models for each pair of low and high $\Gamma$ models ranging from 0 to 100 \%$\Gamma_{\text{low}}$ (henceforth we refer to these models by their $\%\Gamma_{\text{low}}$, where the $\%\Gamma_{\text{high}}$ is complementary). These models correspond to a spatially inhomogeneous mixture of two materials below the resolution of the data, in contrast to the vertically stratified layer model of~\citet{spencer_surfaces_1987}. Note that the combination of variable emissivity and $\%\Gamma_\text{low}$ controls the relative contribution of each thermal inertia component in the same way as two emissivity endmembers, as are used by some other works. Finally, the unconvolved mixture model was convolved with the ALMA synthesized beam and subtracted from the observed emission to generate a thermal residual at the corresponding viewing geometry. We employed the cost function described by~\citet{de_kleer_surface_2021} to obtain comparative $\chi^2$ values from the thermal residuals and determine the best-fit hemispheric thermophysical properties for each observation. To report uncertainties on the derived properties, we employ a standard $\chi^2$ cutoff~\citep[e.g.,][]{hanus_thermophysical_2015,cambioni_heterogeneous_2022}. Namely, for each property we report the set of values satisfying a $\chi^2$ cutoff of $(1 + \sqrt{2/\nu})\chi^2_{\text{min}}$, where $\nu$ is the number of degrees of freedom minus the number of free parameters (typically $\sim$1.05$\chi^2_\text{min}$)

\section{Results \& Discussion} \label{section:results}

We present thermophysical properties derived from six 0.43--0.47 mm (701.9--641.5 GHz) ALMA calibrator observations across all longitudes of Callisto's surface. In Section~\ref{section:GP}, we present disk-averaged brightness temperatures and contextualize our observations among previous unresolved infrared and (sub)millimeter observations of Callisto. We employ a thermophysical linear mixture model to estimate the global $\Gamma$ and emissivity of Callisto's surface (Section~\ref{section:GP}), and subtract the best fitting model from the data to obtain residual maps that hint at the thermal properties of its major geologic features (Section~\ref{section:TM}). In Section~\ref{section:CGPO}, we present several previously unpublished \textit{Galileo} PPR observations of Callisto to further contextualize our results and infer the thermal properties of major geologic features across the surface.

\begin{figure}[t]
\centering
\includegraphics[width=0.45\textwidth]{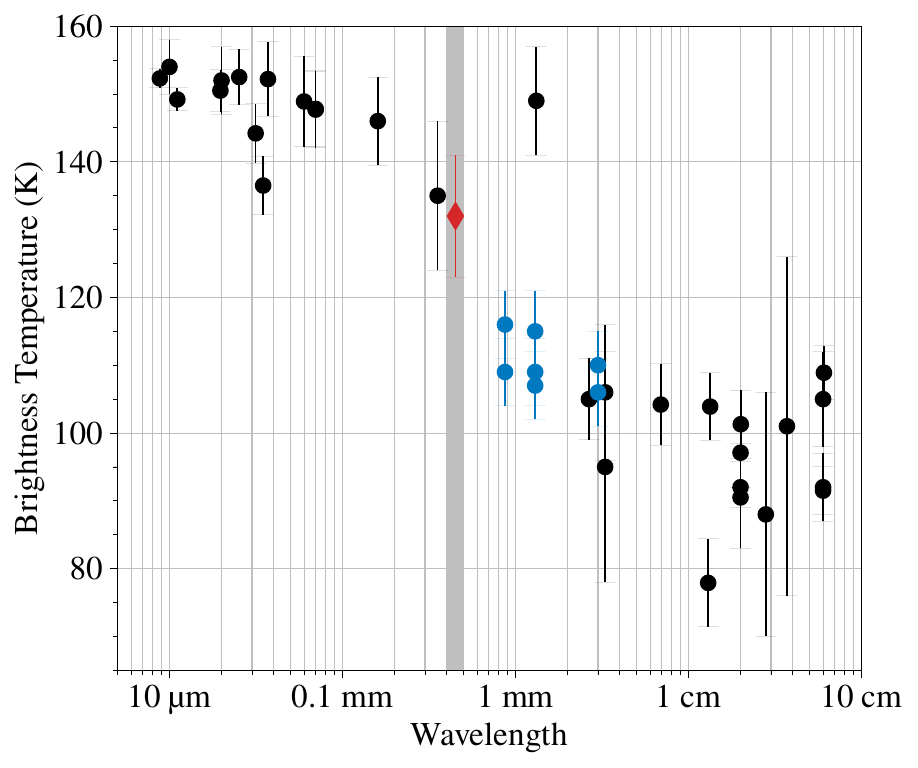}
\caption{Summary of disk-averaged brightness temperature measurements for Callisto as a function of wavelength. Red point represents the average brightness temperature from the present work, with gray shading indicating the wavelength range covered by our ALMA observations. Shown in blue are brightness temperatures obtained from ALMA images~\citep{camarca_thermal_2023,camarca_multifrequency_2025} and black points were retrieved from other previous works~\citep{berge_callisto_1975,de_pater_vla_1984,de_pater_sofia_2021,de_pater_planetary_1989,muhleman_observations_1991,muhleman_precise_1986,pauliny-toth_brightness_1974,pauliny-toth_observations_1977,ulich_millimeter-wavelength_1981,ulich_observations_1976,ulich_planetary_1984,morrison_galilean_1977,muller_far-infrared_2016}.}
\label{fig:bt}
\end{figure}

\subsection{Global properties}\label{section:GP}
\subsubsection{Disk-averaged brightness temperature}\label{section:DIBT}

The derived disk-integrated flux densities range from $39\pm5$ to $80\pm19$ Jy, with corresponding disk-averaged brightness temperatures ranging from $130\pm26$ to $136\pm14$ K (see Table~\ref{table:obsparams}) and a corresponding mean disk-averaged brightness temperature of $132\pm9$ K (corresponding to wavelengths between 0.43 and 0.47 mm). The reported uncertainties incorporate the flux density scale uncertainties reported in Table~\ref{table:fluxcal} and fitting parameter uncertainties. In Figure~\ref{fig:bt}, we summarize these disk-averaged brightness temperatures among other previous measurements obtained by numerous ground-based infrared and long-wavelength facilities, including ALMA~\citep{camarca_thermal_2023,camarca_multifrequency_2025}, the Karl G. Jansky Very Large Array~\citep[VLA;][]{de_pater_vla_1984,muhleman_precise_1986}, Stratospheric Observatory for Infrared Astronomy~\citep[SOFIA;][]{de_pater_sofia_2021}, Infrared Telescope Facility (IRTF) atop Mauna Kea~\citep{de_pater_planetary_1989}, the three-element interferometer at the Owens Valley Radio Observatory~\citep[OVRO;][]{berge_callisto_1975,muhleman_observations_1991}, the 100-m telescope of the Max-Planck-Institut für Radioastronomie~\citep{pauliny-toth_brightness_1974,pauliny-toth_observations_1977}, and the 11- and 12-m reflectors operated by the NRAO at Kitt Peak~\citep{ulich_millimeter-wavelength_1981,ulich_observations_1976,ulich_planetary_1984}. In general, our derived brightness temperatures exhibit strong agreement with past measurements, excepting the measurement of $149\pm8$ K reported by~\citet{ulich_planetary_1984}, which exceeds the neighboring measurements by $\sim$40 K. Importantly, all measurements were obtained by Earth-based observatories and as such, are all day-side measurements. That the brightness temperature appears to drop off between $\sim$50 $\upmu$m and $\sim$5 mm is likely due to a combination of longer wavelengths' sensing of greater depths where the amplitude of the diurnal wave is considerably attenuated and higher $\Gamma$ (more compact) materials at depth. Wavelengths longward of $\sim$5 mm are sensitive to emission originating from below the thermal skin depth ($\sim$tens of cm). As such, the diurnal wave temperature contribution is lost and the curve levels off. A similar decreasing and subsequent leveling off trend is seen for Ganymede, although Io and Europa exhibit slightly different characteristics~\citep[e.g.,][]{de_pater_sofia_2021}.

\begin{figure}[t]
\centering
\includegraphics[width=0.435\textwidth]{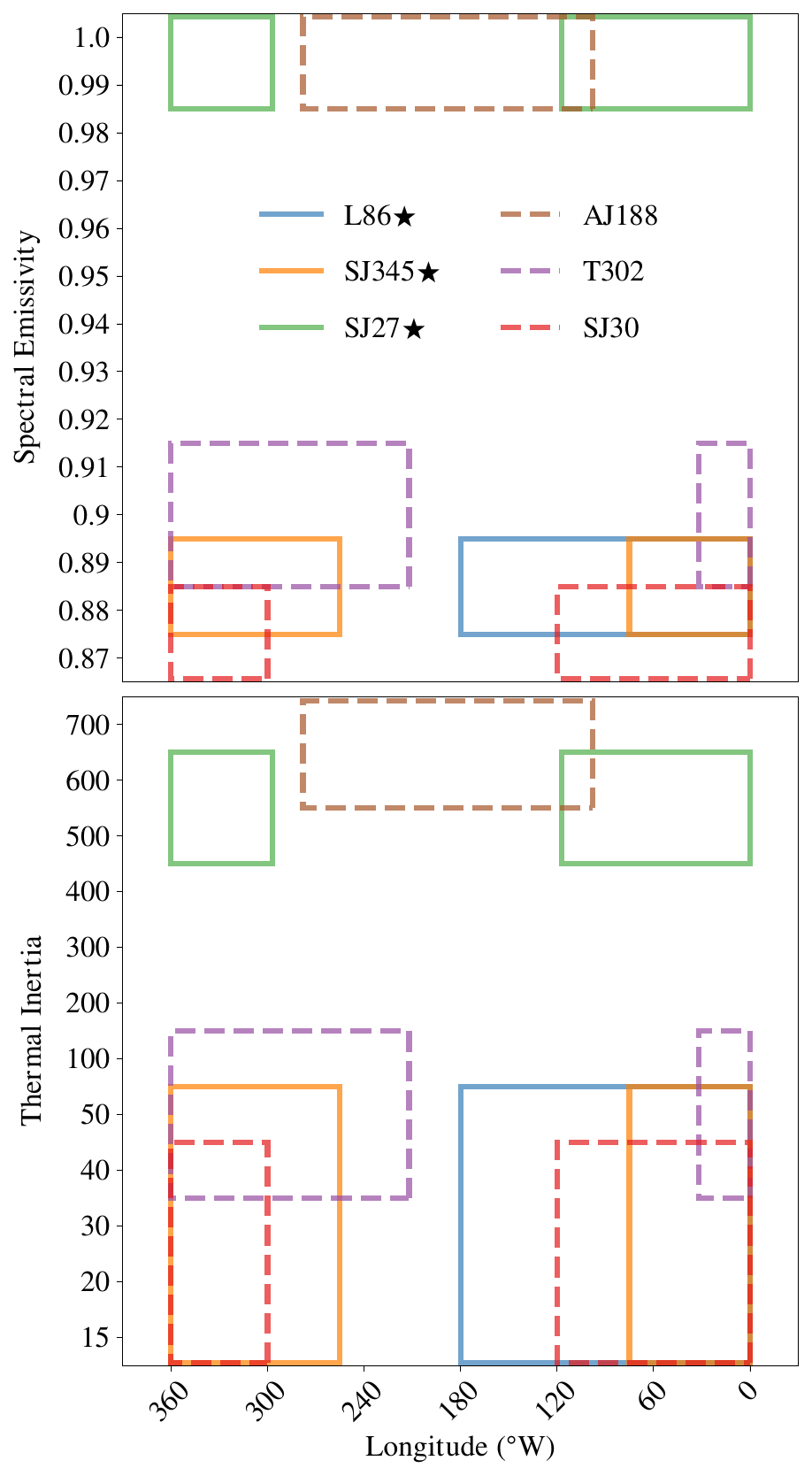}
\caption{Best-fit thermal inertiae and emissivities from single-$\Gamma$ models plotted as a function of longitude across Callisto's surface. Solid and dashed lines correspond to high-quality and low-quality observations, respectively (see Section~\ref{subsection:observations}). The set of $\Gamma$ evaluated ranged from 15 to 2000 but we truncate the plotted range since no best fitting $\Gamma$ exceed 700.}
\label{fig:single_inertias}
\end{figure}

\subsubsection{Constrained single-\texorpdfstring{$\Gamma$}{Gamma} fits}\label{section:CSGF}

We initially constrain the thermophysical model to the use of a single thermal inertia (single-$\Gamma$) in order to directly compare against the quality of the linear mixture models, which yielded optimal fits for 0.87 mm (343 GHz), 1.3 mm (233 GHz), and 3 mm (97 GHz) ALMA observations of Callisto by~\citet{camarca_multifrequency_2025}. Unlike~\citet{camarca_multifrequency_2025}, the thermal inertia was well constrained in single-$\Gamma$ fits for all six observations. We present a summary of the best fitting single-$\Gamma$ global thermal properties in Figure~\ref{fig:single_inertias}. The derived properties fall into two distinct groups, most notably in the emissivity, where SJ27$\bigstar$ and AJ188 have values of 0.99--1 and the others values of 0.87--0.91. For SJ27$\bigstar$ and AJ188, the optimal $\chi^2$ values occur at even greater $\Gamma$ than shown in Figure~\ref{fig:single_inertias} but were limited to $\Gamma$$\sim$500--700 because the emissivities corresponding to higher $\Gamma$ exceeded 1 ($>$1 is non-physical). This, alongside the fact that there does not appear to be any unique shared characteristics between the two observations, suggests that the overall quality of the fits may be poor. Omitting these two observations, we find best fitting emissivities of 0.87--0.91 alongside $\Gamma$$\sim$15--100. All $\chi^2$ values for these models exceed those of their two-$\Gamma$ counterparts (see Section~\ref{section:TGF}).

The derived emissivities exhibit strong agreement with values reported by previous studies. For example,~\citet{camarca_multifrequency_2025} found representative millimeter emissivities of $\sim$0.88--0.95 using lower frequency ALMA observations, and values near unity were reported at 0.355 mm by~\citet{de_pater_planetary_1989} and at 2.66 mm by~\citet{muhleman_observations_1991}. In general, Callisto's millimeter emissivity is higher than those reported for Europa~\citep[$\sim$0.75--0.85; e.g.,][]{thelen_subsurface_2024,trumbo_alma_2018} and Ganymede~\citep[0.75--0.78; e.g.,][]{de_kleer_ganymedes_2021}, likely a consequence of its relatively ice-poor surface since the emissivity of rock is greater than ice at millimeter wavelengths. Our derived $\Gamma$ are remarkably well constrained in comparison with the single-$\Gamma$ values reported by~\citet{camarca_multifrequency_2025}, which ranged from 100 to 2000 with several observations yielding poorly or completely unconstrained $\Gamma$. The depths probed by their lower frequencies (97--343 GHz) are near or below the thermal skin depth, resulting in more subtle temperature variations over each Callisto day and increasingly degenerate thermal models. Higher frequency observations are sensitive to shallower depths, above the thermal skin depth, at which diurnal temperature variations are more pronounced, potentially yielding less degenerate models and tighter constraints on thermal properties.

The thermophysical models employed within this work are global in that each model minimizes $\chi^2$ using a single $\Gamma$-emissivity pair and as such, that pair of global properties represents Callisto's entire visible surface for the observation being fitted. However, we generate unique models for each observation and the set of observations occupy a wide range of sub-observer longitudes. By assigning the $\Gamma$ and emissivity derived from each observation to its corresponding longitudes, we can obtain a low resolution spatial map of derived properties across Callisto's surface. In Figure~\ref{fig:single_inertias}, we present the best fitting emissivities and $\Gamma$ as a function of longitude on Callisto's surface. Such a map is not useful when thermal properties are poorly constrained. The spatial distributions of both the emissivity and thermal inertia are generally uniform across all longitudes, excepting observations SJ27$\bigstar$ and AJ188, as described above. Although the best fitting $\Gamma$ reported by~\citet{camarca_multifrequency_2025} are somewhat poorly constrained, their reported values are also generally similar at each frequency between the leading and trailing hemispheres. Leading/trailing hemispheric differences are generally expected due to exogenic modification processes like micrometeorite bombardment and ice sintering by energetic electrons sweeping past Callisto throughout its orbit, but these expected differences appear to be outweighed by the inability for a single-$\Gamma$ model to adequately fit all parts of Callisto's disk. A more thorough discussion of the role of these processes can be found in Section~\ref{section:TM}.

Following~\citet{thelen_subsurface_2024}, we also explored multiplying the imaginary part of the complex refractive index $\kappa$ (where $\tilde{\eta}=\eta+i\kappa$) by an additional scale factor $c$ between 0.25 and 12.5, corresponding to porous water ice and rocky regolith, respectively. This multiplicative scale factor alters the electrical skin depth via $\delta_\text{elec}=\lambda/(4\pi c\kappa)$, thus modifying the effective absorptivity of the material. This approach was only applied to the single-$\Gamma$ models and is referred to as the single-$\Gamma$$+$variable $\delta_\text{elec}$ model. Unlike the aforementioned authors, we see only negligible improvements in the calculated $\chi^2$ values, with the overall best fitting thermal inertiae remaining nearly constant and the corresponding thermal residuals effectively unaffected between the single-$\Gamma$ and single-$\Gamma$$+$variable $\delta_\text{elec}$ models. Future work may implement a two-$\Gamma$$+$variable $\delta_\text{elec}$ model, although~\citet{camarca_multifrequency_2025} found that the two-$\Gamma$ and single-$\Gamma$$+$variable $\delta_\text{elec}$ models improved upon the single-$\Gamma$ fits in similar ways. As such, an increasingly complex two-$\Gamma$$+$variable $\delta_\text{elec}$ model is likely to yield only minor improvements to the aforementioned models.

\begin{figure*}[t]
\centering
\includegraphics[width=0.99\textwidth]{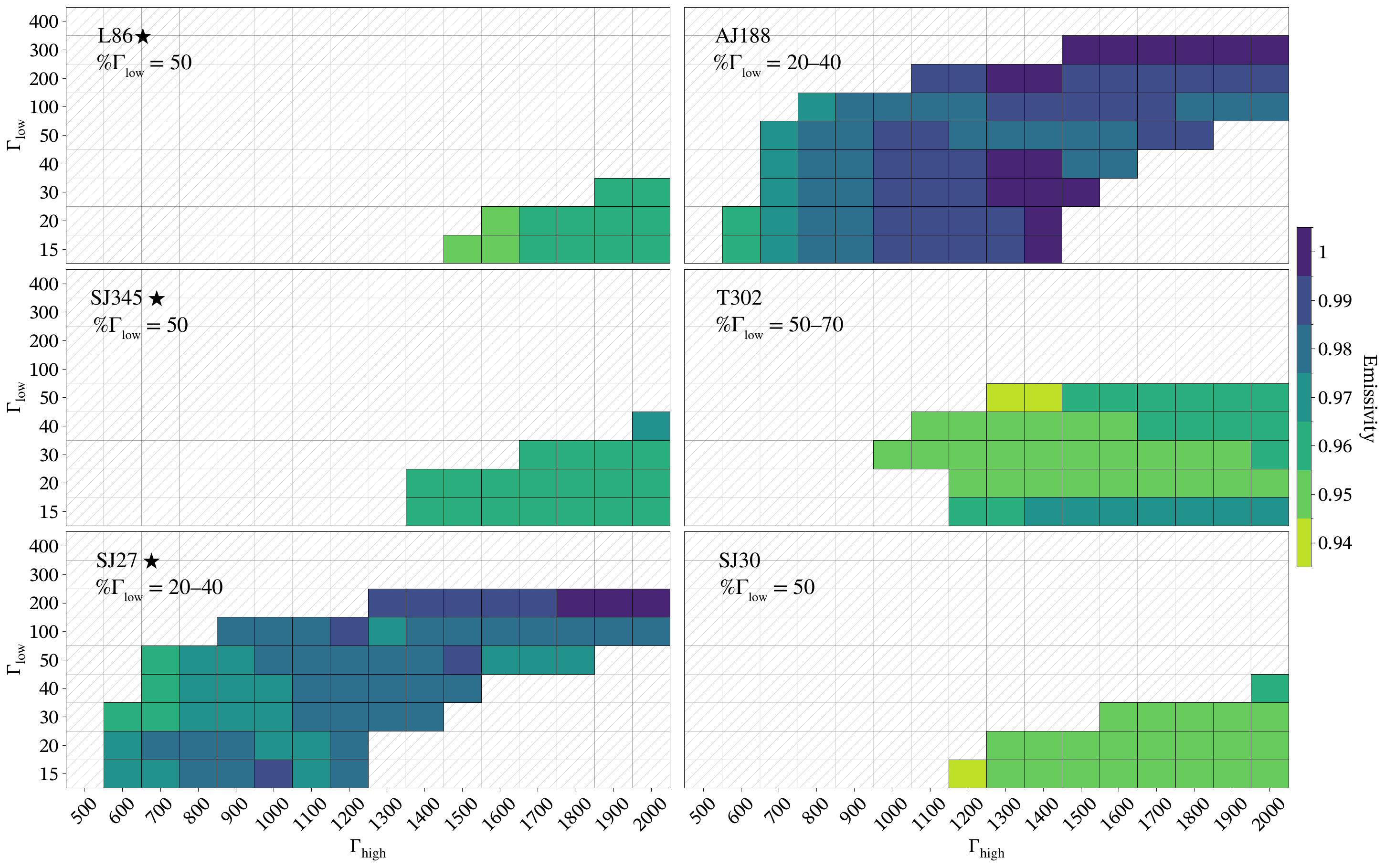}
\caption{Best fitting two-$\Gamma$ derived thermal inertiae and emissivities. Hashed squares correspond to high-low $\Gamma$ pairs whose optimal $\chi^2$ values exceed the $\chi^2$ cutoff described in Section~\ref{section:model} (typically $\sim$1.05$\chi^2_{\text{min}}$). Each subplot corresponds to a single observation whose code is in its top-left corner and all six subplots share the same color scale. Observations previously deemed high- and low-quality are separated into the left and right columns, respectively.}
\label{fig:emiss_grid}
\end{figure*}

\subsubsection{Two-\texorpdfstring{$\Gamma$}{Gamma} fits}\label{section:TGF}

The tendency for high and low $\Gamma$ models to underestimate Callisto's emission near its center and at the limb, respectively, is ubiquitous across our six observations, and has been identified previously by, e.g.,~\citet{spencer_surfaces_1987}. In the case of the single-$\Gamma$ models, the higher $\Gamma$ models offered a better fit to the data than the lower thermal inertia models in general. However, their corresponding thermal residuals suggest that a better fit may be obtained by a two-$\Gamma$ model that combines the center fit of the low thermal inertia with the limb fit of the high $\Gamma$~\citep[see Figure 2 of][for an example of this concept]{camarca_thermal_2023}. As such, we generate a grid of thermal models -- for each high and low thermal inertia pair, we linearly combine their unconvolved models in relative increments of 10\%, resulting in 11 mixtures for each pair. The linearly mixed unconvolved model is then convolved with the ALMA beam to obtain a model of the observed emission (see Section~\ref{section:model} for more details). Figure~\ref{fig:emiss_grid} summarizes the best-fitting thermal properties of these two-$\Gamma$ models, with the color scale representing emissivity and $\%\,\Gamma_{\text{low}}$ printed on the top left of each panel (where $\%\,\Gamma_{\text{high}}=100-\%\,\Gamma_{\text{low}}$) to highlight grid spanning trends.

As in the case of the single-$\Gamma$ models, the two-$\Gamma$ results fall into two distinct categories: SJ27$\bigstar$ and AJ188, and the others. The latter group is generally well fit by $\Gamma_{\text{low}}$$\sim$15--40 and $\Gamma_{\text{high}}$$\sim$1200--2000, with $\%\,\Gamma_{\text{low}}$$\sim$50--60 and emissivities between 0.94 and 0.97, although the fits differ slightly between observations. These optimal mixtures of very low and very high thermal inertiae are qualitatively expected based on the aforementioned inability for high and low thermal inertiae to match emission near Callisto's center and limb, respectively, on their own. The four observations collectively span nearly all longitudes across Callisto's surface, making the relative uniformity of their best fitting thermal properties notable, though local trends in the residuals will provide additional context (see Section~\ref{section:TM} for a thorough discussion). Observations SJ27$\bigstar$ and AJ188 exhibit considerably higher emissivities than the others (as was the case for the single-$\Gamma$ models) and nearly identical diagonal trends in the thermal inertia grid of Figure~\ref{fig:emiss_grid}. The $\%\,\Gamma_{\text{low}}$ values tend to increase toward the top-right corner of the two thermal inertia grids, which in combination with the positive slope of their best fitting $\Gamma$ values, indicates a preference toward moderate to high thermal inertiae. Specifically, at low $\Gamma_{\text{low}}$ values, optimal fits occur for low $\%\,\Gamma_{\text{low}}$ and $\Gamma_{\text{high}}$ of $\sim$600--1200 (i.e., thermal inertiae of $\sim$600--1200 dominate), whereas at high $\Gamma_{\text{low}}$, the $\%\,\Gamma_{\text{low}}$ increases to $\sim$40. Despite these apparent trends toward high $\Gamma$ values, two-$\Gamma$ models consistently exhibit better $\chi^2$ values than their single-$\Gamma$ counterparts, suggesting a maintained tradeoff between high and low $\Gamma$ fits even for models which exhibit potentially anomalous emissivities~\citep[for a visual demonstration of this tradeoff, see Figure 2 of][]{camarca_thermal_2023}. Since the two-$\Gamma$ models exhibited consistently lower $\chi^2$ values compared to the single-$\Gamma$ models, we adopt the former for subsequent analysis of the thermal residuals in Section~\ref{section:TM}. That said, the anomalous behavior of SJ27$\bigstar$ and AJ188 in both the single- and two-$\Gamma$ fits are consistent with our suspicion that the data quality may be reduced for those observations. The observations are included for completeness but should be interpreted with caution.

\begin{figure*}[t]
\centering
\includegraphics[width=0.78\textwidth]{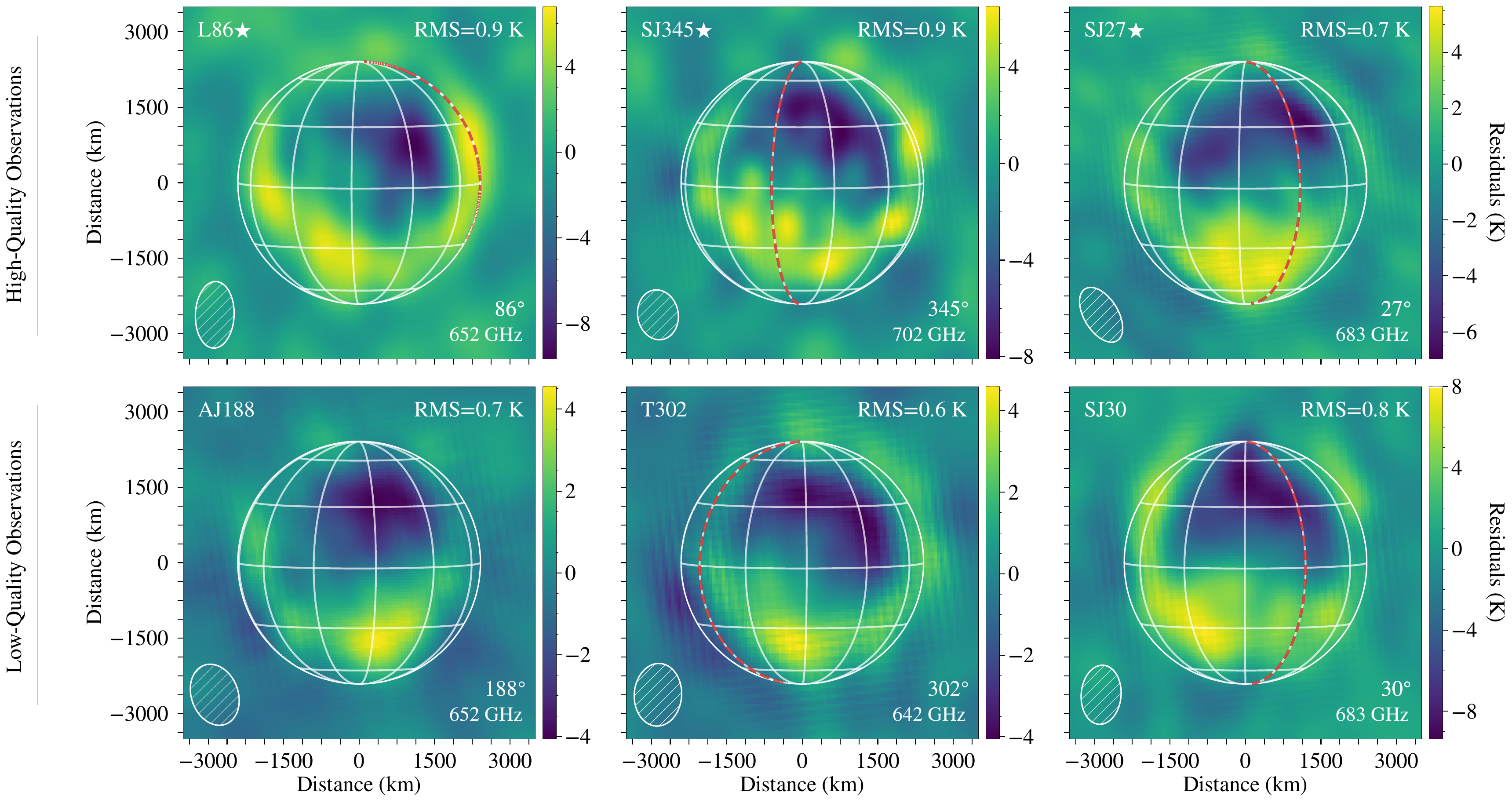}
\caption{Residuals (ALMA data minus model) generated using the best fitting thermophysical models described in Section~\ref{section:results}. The RMS (measured from a non source region of the non primary beam-corrected image product) is shown in the top right of each subplot. Each subplot uses a separate color scale to enhance the visibility of structure in the residuals.}
\label{fig:TPM_subtract}
\end{figure*}

\begin{figure*}[t]
\centering
\includegraphics[width=1\textwidth]{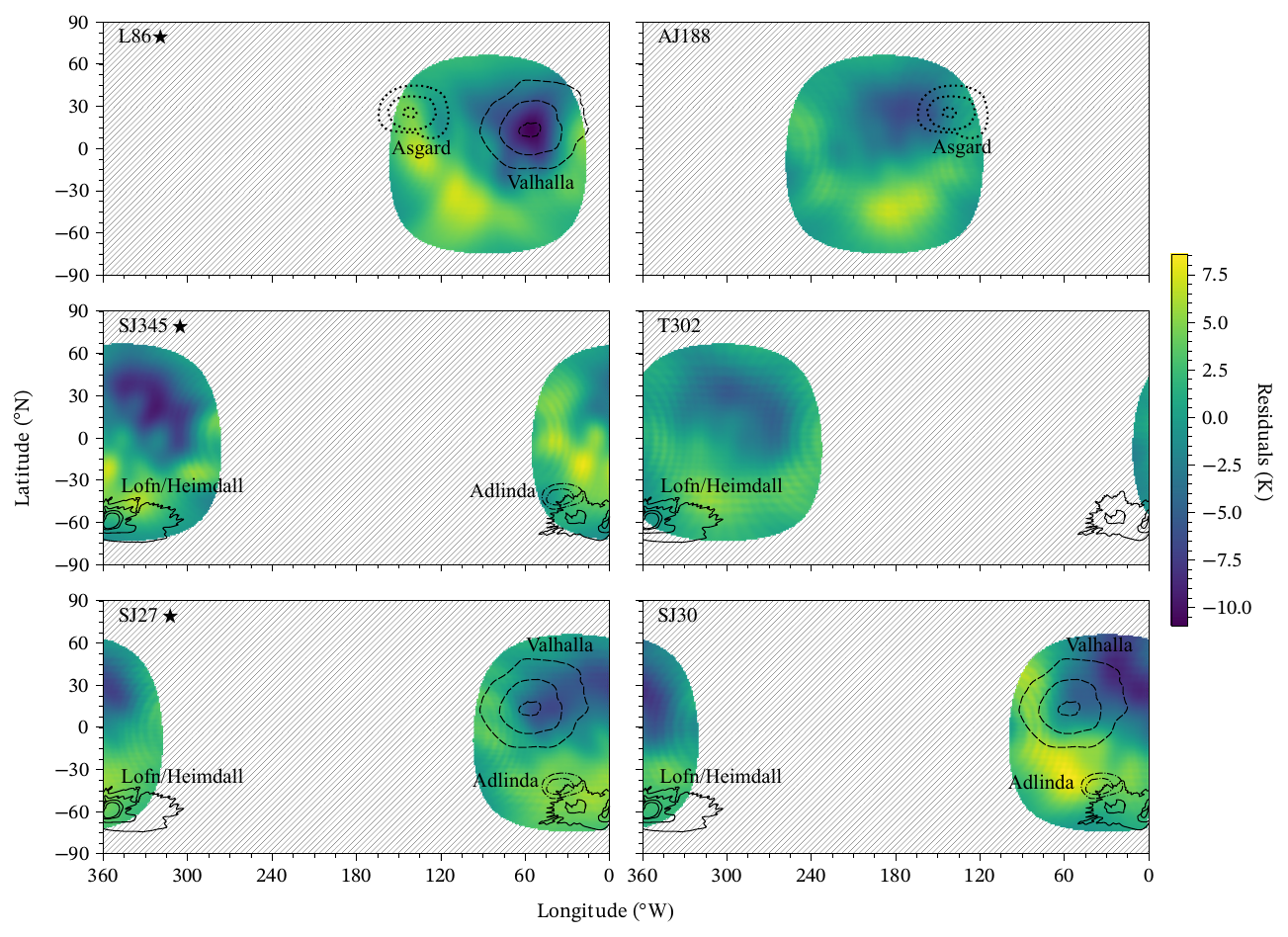}
\caption{Deprojected thermal residuals for the best fitting two-$\Gamma$ models with contours~\citep{greeley_galileo_2000} identifying overlapping major geologic features. Valhalla (dashed), Asgard (dotted), Adlinda (dash-dotted), and Lofn/Heimdall (solid) are indicated by contours and labels}. Each residual includes emission angles of $\leq$70$^\circ$ to exclude emission presumed altered by edge effects, and all six subplots use the same color scale. The code for each observation is in its top-left corner, and high- and low-quality observations are shown in the left and right columns, respectively.
\label{fig:deprojections}
\end{figure*}

The implication that Callisto's surface may be composed of loose, finely particulated material ($\Gamma$$\sim$15--40), in combination with competent (bed)rock-like material ($\Gamma$$\sim$1200--2000) in comparable proportions suggests notable differences in regolith composition and/or formation mechanisms compared with similar inner solar system bodies. In general, one might expect an old surface such as that of Callisto to be comprised primarily of finely particulated, pulverized regolith with low thermal inertia. For instance, \textit{Lunar Reconnaissance Orbiter} (LRO) Diviner Lunar Radiometer Experiment revealed low global thermal inertiae of $\sim$55$\pm$2 for the Moon~\citep{hayne_global_2017}, similar to that of Mars~\citep{putzig_global_2005} and main-belt asteroids~\citep{delbo_thermal_2009,spencer_rough-surface_1990,capria_vesta_2014}. However, we find that a non-negligible high thermal inertia component must be present on Callisto's surface, consistent with lower frequency ALMA data~\citep{camarca_thermal_2023,camarca_multifrequency_2025}, and indeed, there appears to be a trend of increasing thermal inertiae with distance from Jupiter (see, for instance, Table 4 of~\citealp{thelen_subsurface_2024} and Figure 6 of~\citealp{camarca_thermal_2023}). While it is challenging to deduce which regolith forming mechanisms may be responsible for this trend, we note that high-resolution \textit{Galileo} data revealed exposed ice blocks on Callisto's surface~\citep[e.g.,][]{chuang_large_2000} that may be responsible. These blocks could be relatively unaltered compared with ice on other icy moons, which are subject to particle bombardment and subsequently diminished thermal inertiae via impact gardening. Additionally, likely contributing to a lesser extent, the Galileans' water ice phase tends toward crystalline (from amorphous) with increasing distance from Jupiter~\citep[e.g.,][]{ferrari_thermal_2018}. Since crystalline water ice has a considerably higher thermal inertia than amorphous water ice~\citep{ferrari_low_2016}, one may expect the outer Galileans to have higher global thermal inertiae even with comparable ice fractions and porosities.

\subsection{Residual thermal maps}\label{section:TM}

By subtracting the best fitting thermophysical models from each observation, thermal residuals are obtained which reveal regions not well fit by the model, indicative of surface properties that differ locally from the global average. All localized features seen in thermal residuals are well above the noise level (>$5\sigma$, where $\sigma$ is the RMS noise measured from a non source region of the non primary beam-corrected image product) and fairly robust to changes in model parameters (i.e., residuals are similar across best fitting $\Gamma_{\text{low}}$ and $\Gamma_{\text{high}}$ pairs). Poorly fit regions are thus interpreted as regions of altered thermal properties and linked directly to geologic features when possible. Figure~\ref{fig:TPM_subtract} shows the thermal residuals obtained by subtracting each best-fitting model from its corresponding ALMA image. To more readily identify common features between observations, we deproject their corresponding residuals into cylindrical coordinates in Figure~\ref{fig:deprojections} using a shared color scale and label major geologic surface features. 

Anomalously warm or cool regions within the residuals, if not an artifact generated during deconvolution, indicate localized regions of altered thermal properties. For example, impact features on Ganymede~\citep{de_kleer_ganymedes_2021,brown_microwave_2023}, Europa~\citep{trumbo_alma_2017,trumbo_alma_2018,thelen_subsurface_2024}, Titan~\citep{janssen_titans_2016}, and Rhea~\citep{bonnefoy_rheas_2020,howett_thermophysical_2014} were co-located with cool residuals at infrared through microwave wavelengths, likely resulting from locally high thermal inertiae due to a higher fraction of fresh ice in these regions that is exposed to the surface by the impact. Alternatively, global trends may be induced by exogenic sculpting mechanisms like high energy electron sintering~\citep[e.g.,][]{howett_high-amplitude_2011,howett_pacman_2012} and micrometeorite bombardment~\citep[e.g.,][]{ries_large-scale_2015}. Altogether, an analysis of the residuals complements the derived thermal properties through deviations from the thermophysical model, which does not consider endo- or exogenic modification mechanisms or local variations in thermal properties induced by geologic surface features.

\subsubsection{Valhalla and Asgard}

Valhalla is the largest multi-ring impact basin in the solar system, spanning 3800 km (including outer troughs) or $\sim$16\% of Callisto's surface area. Despite lacking a clearly definable rim, the basin is delineated into a smooth central region, an inner ridge and trough zone, and an outer trough zone~\citep{mckinnon_evolution_1980,schenk_geology_1995,greeley_galileo_2000}, shown as separate contours in Figure~\ref{fig:deprojections}. In the context of our ALMA observations, Valhalla boasts the best coverage of any major surface feature. Across all three observations with considerable coverage by Valhalla (SJ27$\bigstar$, SJ30, and L86$\bigstar$), cool spots co-locate with its central region at $\sim$18$^\circ$N and $\sim$57$^\circ$W. The residuals range from --11 to --5 K for L86$\bigstar$/SJ30 and SJ27$\bigstar$, respectively, compared with values of around --5, --2, and --4 K, for observations at 3, 1.3, and 0.87 mm, respectively~\citep{camarca_multifrequency_2025}. The depths probed by our wavelengths experience more significant diurnal temperature variations than those probed by the longer wavelengths of~\citet{camarca_multifrequency_2025} (recall that the diurnal wave amplitude falls off like $e^{-z/\delta_\text{therm}}$ and that the depth probed is given by the electrical skin depth $\delta_\text{elec}=\lambda/(4\pi \kappa)$; see Section~\ref{section:DIBT} for more discussion), resulting in increased residual magnitudes. As such, our deepened residuals at Valhalla compared to those reported by~\citet{camarca_multifrequency_2025} is expected and likely part of a broader trend (see Section~\ref{section:CGPO} for further exploration). That said, the magnitude of thermal residuals will depend on the local time of day at each region on the disk even when the thermal model accounts for the diurnal wave -- residuals' magnitudes should be interpreted with this in mind.

Asgard is the second largest multi-ring impact basin on Callisto, located around 30$^\circ$N, 142$^\circ$W, the majority of which is dotted with bright, fresh craters like Doh, Burr, and Utgard and their ejecta. Our spatial coverage of Asgard is limited to AJ188 and L86$\bigstar$, and there is no obvious overlap between Asgard and a thermal feature. However, a cool region of --4 K (4--5$\sigma$) lies immediately westward of Asgard, with a portion of the feature overlapping with Asgard's westernmost boundary. A similar cool region was identified within 97 and 345 GHz ALMA imagery by~\citet{camarca_multifrequency_2025}, who tentatively attributed the feature to the small, bright crater Nirkes lying directly west of Asgard. As is the case with the residuals from~\citet{camarca_multifrequency_2025}, the spatial extent of the cool region is approximately the same size as the ALMA beam and thus may indeed result from a small localized feature such as Nirkes.

\subsubsection{Adlinda/Heimdall/Lofn Complex}

The cold crater phenomenon described in Section~\ref{section:TM}, which holds that the high thermal inertia materials exposed by impacts tend to maintain lower daytime temperatures than their surroundings, does not hold for all impact features universally. For instance, the large impact feature Tashmetum of Ganymede did not appear anomalous compared to its surroundings despite being classified as fresh alongside the craters Tros and Osiris, which each exhibited clear thermal anomalies~\citep{de_kleer_ganymedes_2021,collins_global_2014}. The impact features Adlinda, Lofn, and Heimdall spanning the southern quadrant of Callisto's subjovian hemisphere exhibit similar exceptions to the cold crater phenomenon in observations SJ345$\bigstar$, SJ27$\bigstar$, T302, and SJ30. In contrast, lower frequency ALMA images at 233 and 343 GHz exhibited a clear cool anomaly of magnitude --3 to --4 K co-located with the complex~\citep{camarca_multifrequency_2025}. In the cases of SJ345$\bigstar$ and T302, the muted residuals co-located with the Adlinda/Lofn/Heimdall complex may be due to geometric foreshortening induced by their proximity to Callisto's limb (in both cases, the spatial extent of the complex is comparable to the extent of the ALMA beam). However, the trend remains seen in SJ27$\bigstar$ and SJ30, whose observational centers lie within $\sim$60$^\circ$ of Adlinda and Lofn, suggesting that the muted temperatures may be physical as opposed to artifacts of the data.

\subsubsection{Other notable features}

In SJ27$\bigstar$ and SJ30, the cool spot associated with Valhalla extends eastward (toward $\sim$350$^\circ$) in an upside-down ``U'' shape with larger residuals (--8 K) than Valhalla itself (see SJ30 in Figure~\ref{fig:TPM_subtract} for a clear example). Although such a feature is unexpected based on overlapping observations from~\citet{camarca_multifrequency_2025} and the lack of geologic features immediately east of Valhalla, a continuation of the trend further eastward (toward $\sim$270$^\circ$) is seen in SJ345$\bigstar$ and T302.

Similarly intriguing are the warm residuals surrounding several of the cool features outlined above (e.g., the ``L''-shaped warm spot southwest of Valhalla in L86$\bigstar$, SJ27$\bigstar$, and SJ30, the warm region south of Asgard in AJ188, etc.). Each of these warm regions correspond to observations that contain at least one significant, cool region correlated with a surface feature. As such, the spectral emissivities for the corresponding observations were likely depressed by the cool spots in order to minimize the overall $\chi^2$ values, resulting in $>$0 K residuals in regions outside the cool features (including introducing anomalously warm regions and lower magnitude cool regions). In other words, while low spectral emissivities may minimize global $\chi^2$ values for models with deeply negative residuals (e.g., co-located with an impact crater), those same low emissivities will force the model lower elsewhere (e.g., across cratered plains), yielding artificially positive residuals outside the anomaly. To minimize this bias, one could mask out the regions expected to correspond to thermal anomalies and optimize the spectral emissivity to fit the ``normal'' regions of Callisto's surface, yielding more representative thermal residuals and cool region magnitudes. However, the relatively large resolution elements of ALMA ``smear out'' thermal anomalies and as such, one would need to mask out a considerable fraction of the residuals to achieve the desired effect. Instead, we simply emphasize that the thermal residuals of images that have significant cool regions may be systematically warmer as a result of the bias.

\subsection{Comparison with Galileo PPR observations}\label{section:CGPO}

The Band 9 ALMA observations presented above uniquely bridge the gap between lower frequency ALMA data~\citep{camarca_multifrequency_2025} and thermal infrared data in depth probed, though few spatially resolved observations of Callisto in the thermal infrared have been published previously. To contextualize our ALMA observations amongst thermal infrared data, we present previously unpublished thermal observations obtained by the \textit{Galileo} Photopolarimeter-Radiometer~\citep[PPR;][]{russell_galileo_1992,travis_galileo_2000} in radiometer mode between 1996 November 4 and 2001 May 25. These observations represent the first PPR radiometer data to be published for Callisto and complement the spectral coverage of our ALMA observations with shorter wavelengths, which probe more shallow subsurface depths. In particular, we analyzed 21 observations distributed across orbits C3, C10, C21, C30, E6, E11, and G7 using channels A and C, with central wavelengths of 16.8 $\upmu$m and 27.5 $\upmu$m, respectively. The angular resolution for PPR is fixed by a circular field stop subtending 2.5 mrad. The distance between the spacecraft and the field-of-view intercept point ranged from 0.05 to 9.6 R$_\text{J}$, resulting in spatial resolutions of 9 to 1680 km. During preflight calibration, the S/N performance for PPR radiometry was estimated to be $\sim$30 for a 130 K blackbody calibrator device at 16.8 $\upmu$m. Altogether, the observations span $\sim$35\% of Callisto's surface (with $\sim$32\% being usable after masking pixels with robustness parameters $\mathcal{R}=0$; see Section~\ref{section:PPRinertiae} for details).

\begin{figure}[b]
\centering
\includegraphics[width=0.49\textwidth]{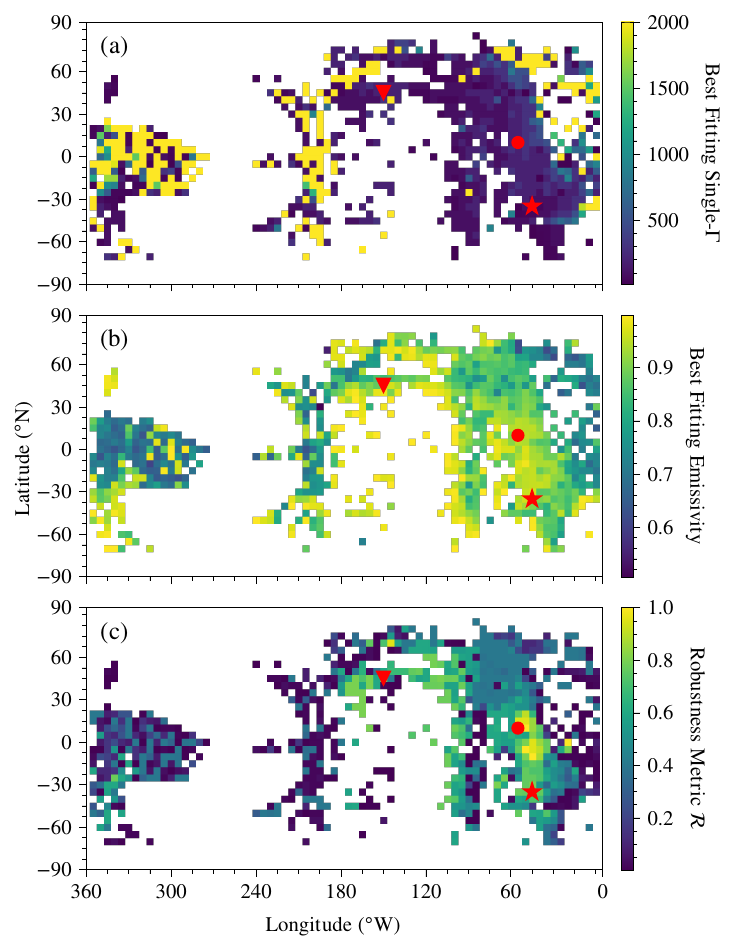}
\caption{Maps of PPR observations. (a) Overall best fitting thermal inertia for each 5$^\circ$ square bin, obtained by fitting the thermophysical model described in Section~\ref{section:model} to the PPR observations described in Section~\ref{section:CGPO}. (b) Corresponding best fitting spectral emissivities. (c) Robustness parameter $\mathcal{R}$ described in Appendix~\ref{section:robustness}. The (c) subplot may be used as a proxy for the ability for the model to constrain the thermal inertia using the available PPR observations. Each subplot uses its own color scale. The red circle, triangle, and star correspond to Valhalla, Asgard, and cratered plains, respectively. In all maps, pixels corresponding to robustness parameters $\mathcal{R}=0$ are masked out.}
\label{fig:PPRmaps}
\end{figure}

\begin{figure*}[t]
\centering
\includegraphics[width=1\textwidth]{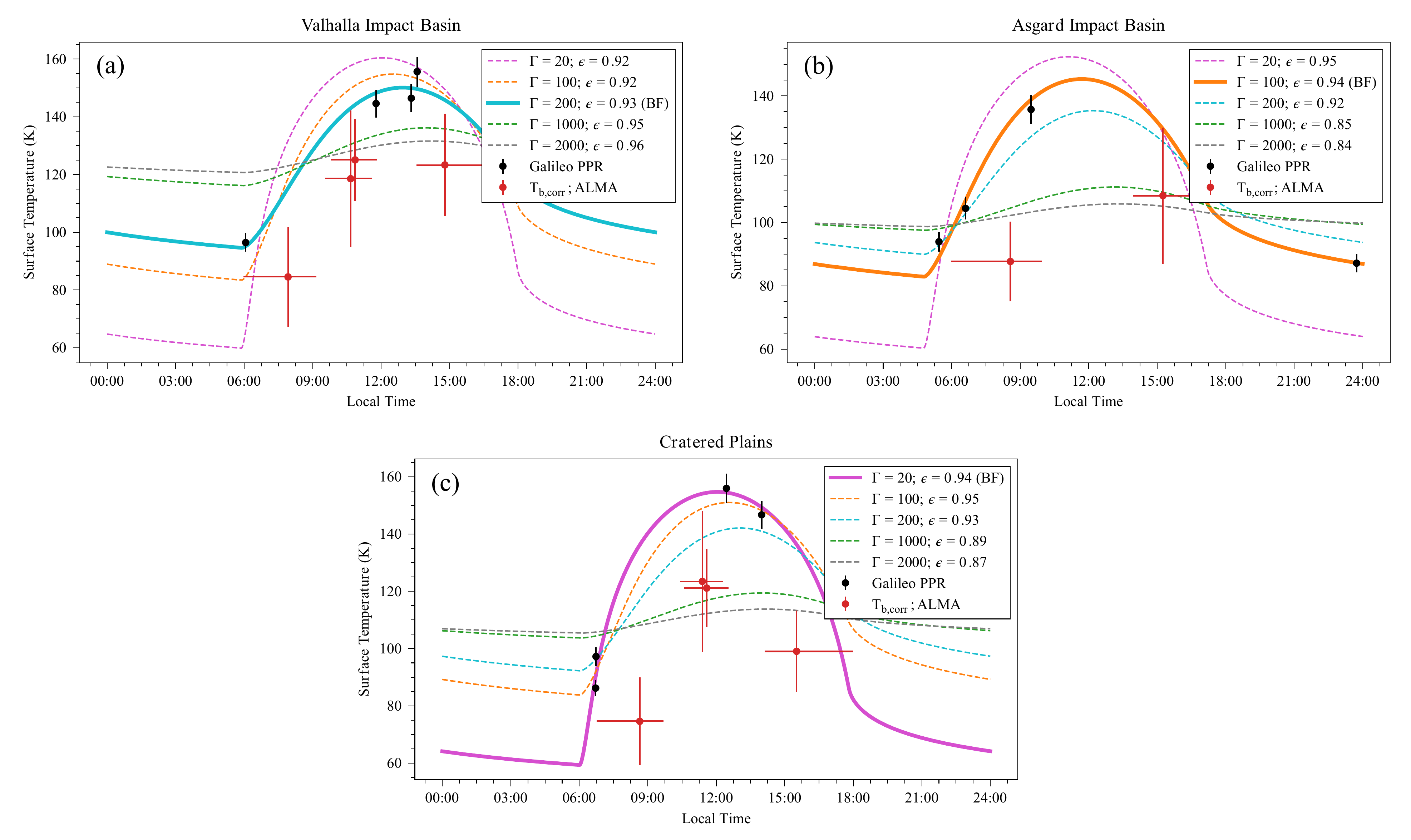}
\caption{Surface brightness temperature (i.e., surface temperature multiplied by spectral emissivity) as a function of time of day for representative regions (Valhalla at 12.5 $^\circ$N and 57.5 $^\circ$W, Asgard at 47.5 $^\circ$N and 152.5 $^\circ$W, and relatively featureless cratered plains at --32.5 $^\circ$N and 47.5 $^\circ$W) on Callisto's surface to emphasize thermal anomalies. Each subplot depicts diurnal curves (lines) across several thermal inertiae values fitted to PPR data (black points) using varying spectral emissivity values. Limb-darkening corrected ALMA subsurface brightness temperatures (i.e., brightness temperatures that receive contributions from subsurface emission) sampled at the above coordinates are shown as red points for reference, where horizontal error bars indicate the extent of the ALMA beam projected onto the surface; the ALMA data are not included in the fits since they probe different depths than the PPR surface temperature data. Note that since the subsurface emission that ALMA senses originates nearer to the thermal skin depth, diurnal temperature variations will be more muted even with the same thermal inertia. The best fitting curve at each location is made solid and bold, whereas other curves are dashed.}
\label{fig:multiwave}
\end{figure*}

The PPR data were obtained directly from the Outer Planets Icy Satellites Archive Page\footnote{\href{https://atmos.nmsu.edu/data_and_services/atmospheres_data/JUPITER/icy_satellites.html}{https://atmos.nmsu.edu/data\_and\_services/atmospheres\_data/ JUPITER/icy\_satellites.html}} hosted on the Planetary Atmospheres Node of the Planetary Data System (PDS). We employ the reduced data record (RDR) version of the data, which were generated by the PPR team using the standard reduction pipeline and deboomed (corrected for obscuration by the antenna boom). The RDR data are provided as apparent brightness temperatures at each timestep alongside the latitudes and longitudes of the field of view intercept points on Callisto's surface. To generate the thermal maps, we gridded each data point at its intercept point as a circle whose diameter equals the spatial resolution. For each observation, we assumed uniform brightness temperature across each data point, averaged overlapping regions, and binned the data into 5$^\circ$ squares. Doing so resulted in a list of brightness temperatures with corresponding local times and uncertainties for each bin across the surface. Each element of this list represented an individual PPR observation (i.e., averaging across observations was not performed), allowing the diurnal curve to be constrained to an extent determined by the temporal coverage of brightness temperature measurements in the corresponding bin. In general, constraining a diurnal curve is challenging with only one or two brightness temperature measurements, even at different times of day. To effectively visualize regions across the maps which offer more robust constraints on the inferred thermal inertiae and emissivities, we introduce a ``robustness metric'' $\mathcal{R}$ which reflects both the number of measurements and the width of thermal inertiae $w$ that fall within twice the minimum $\chi^2$ (e.g., $w=385$ for well fitting thermal inertiae of 15 to 400) at each bin. Values near null are considered least robust, whereas values near unity are most robust. The robustness metric $\mathcal{R}$ is defined, described in depth, and depicted alongside similar metrics like $\chi^2$ in Appendix~\ref{section:robustness}.

\subsubsection{\textit{Galileo} PPR thermophysical modeling}

We employ the single-$\Gamma$ thermophysical model described in Section~\ref{section:model} for simplicity, with emission arising only from the immediate surface for infrared emission. Since the observed brightness temperature depends upon observing geometries specific to each PPR observation, a single thermal model cannot simultaneously fit multiple points without normalization of some kind. To normalize the observations to a common heliocentric distance for comparison purposes, we convert the brightness temperature to flux density using the Planck equation, scale that flux density using the relationship between two fluxes ($F_1$ and $F_2$) of the same source observed at different distances ($r_1$ and $r_2$) $F_1/F_2 = (r_2/r_1)^2$, and invert the Planck function to convert back to brightness temperature. Finally, the varying sub-solar and sub-observer latitude and longitude relationships can readily be converted to local time, which is used as the independent variable for the brightness temperature points to which the diurnal curves are fit.

At each bin across Callisto's surface with at least two PPR observations, we generated diurnal curves for each thermal inertia described in Section~\ref{section:model}. Finally, we fit each diurnal curve to the PPR brightness temperature points within the bin using the spectral emissivity as a scaling factor on the diurnal curves between 0 and 1. Figure~\ref{fig:PPRmaps} presents the best fitting thermal inertia and emissivity at each bin alongside the number of PPR observations and the width of thermal inertiae that satisfy twice the minimum $\chi^2$, as described above (i.e., the robustness metric). Figure~\ref{fig:multiwave} presents the PPR measurements and best fitting diurnal curves within each of three regions of interest: Valhalla (12.5$^\circ$N, 57.5$^\circ$W), Asgard (27.5$^\circ$N, 142.5$^\circ$W), and a region relatively free of geologic features characterized by~\citet{greeley_galileo_2000} as cratered plains (--7.5$^\circ$N, 107.5$^\circ$W). Relevant ALMA brightness temperatures are also shown as a reference but were not included in the fits due to their considerably larger electrical skin depths.

\subsubsection{Thermal inertiae and emissivity spatial distributions}\label{section:PPRinertiae}

Figure~\ref{fig:PPRmaps} presents the best fitting thermal inertia, spectral emissivity, and robustness metric distributions across Callisto's surface. A considerable portion of the available PPR data are not included in these maps due to (1) coverage by only one PPR observation and/or (2) poor fits, indicated by optimal spectral emissivities taking on non-physical values of $>$1. We masked out all pixels corresponding to robustness parameters $\mathcal{R}=0$ since the thermal inertiae for those pixels were completely unconstrained. The fits corresponding to the remaining bins span the entire range of tested thermal inertiae and spectral emissivities between 0.50 and 0.99, with thermal inertiae $>$1500 typically corresponding to robustness metrics of 0 to 0.5. The sparse thermal inertia/emissivity coverage at various regions across the surface makes identifying global trends challenging. However, there exists a band of well constrained models near 60$^\circ$W spanning nearly all latitudes. The band exhibits generally increasing thermal inertia from $\sim$105$^\circ$W toward $\sim$30$^\circ$W, as well as heightened spectral emissivities centered on Valhalla and decreasing in all directions. Despite sparse coverage between $\sim$90$^\circ$W and $\sim$195$^\circ$W, there appears to be heightened emissivities in the entire region which decreases at decreasing longitudes. 

Figure~\ref{fig:multiwave} highlights three regions of interest from the thermal maps of Figure~\ref{fig:PPRmaps}: Valhalla, Asgard, and a region characterized as cratered plains. Whereas the relatively sparse coverage of the thermal maps precludes a robust comparison between these regions, Figure~\ref{fig:multiwave} clearly demonstrates that (1) the three regions have roughly the same spectral emissivity at these infrared wavelengths, and (2) Valhalla and Asgard have considerably higher best-fitting thermal inertiae than the cratered plains region, as would be expected for plains containing fluffier and/or smaller-grained regolith. In the case of modeling a single thermal image (as with the ALMA observations), localized regions of high thermal inertia and low emissivity can both mute the local thermal emission and as such, the two are somewhat degenerate. However, the thermal inertia and emissivity affect the underlying diurnal curve in different ways, where the thermal inertia affects the amplitude of diurnal temperature variations and the emissivity scales the entire curve. Therefore, with observations across multiple times of day at the same location, the two thermal properties can be separated. In other words, we are only able to conclude that Valhalla and Asgard have higher thermal inertiae rather than lower emissivities because we have \textit{Galileo} PPR temperatures across multiple times of day at those locations. That said, spectral emissivities may vary with wavelength and thus may still be lower at Valhalla and Asgard at ALMA wavelengths despite their nearly constant values at infrared wavelengths.

\section{Conclusion}\label{section:conclusion}

We present six thermal images of Callisto at submillimeter wavelengths between 0.43 and 0.47 mm (701.9 GHz and 641.5 GHz, respectively) with coverage across nearly all longitudes of its surface. Using a simple single thermal inertia model, we found best fitting thermal inertiae of $\Gamma$$\sim$15--100 (J m$^{-2}$ K$^{-1}$ s$^{-1/2}$) and spectral emissivities of 0.87--0.91. However, a more complex linear mixture model using two thermal inertiae concurrently yielded considerably better fits. Our best fitting two-$\Gamma$ models adopted $\Gamma_\text{low}$$\sim$15--40 and $\Gamma_\text{high}$$\sim$1200--2000 for $\%\Gamma_\text{low}$$\sim$50--60 and spectral emissivities of 0.94--0.97. These values are nearly identical to those identified by~\citet{camarca_multifrequency_2025} using ALMA data at lower frequencies, although our single-$\Gamma$ fits are markedly better constrained than theirs, likely due to increasingly muted diurnal temperature variations at probed depths approaching the thermal skin depth. By directly comparing the corresponding thermal residual maps against an albedo map of Callisto's surface, we additionally identify a cold temperature anomaly of $\sim$5 K co-located with the Valhalla impact basin and a similarly cool anomaly of $\sim$4 K near the Asgard impact basin, indicating localized high thermal inertia likely resulting from excavation of surface material by impacts. In general, this effect is only expected on icy surfaces, which generally exhibit increasing rock-to-ice ratios with depth. Future work that includes simultaneous fits across spatial orientations and frequencies will enable us to more directly constrain the depth dependence of these thermal properties.

We additionally present several previously unpublished \textit{Galileo} Photopolarimeter-Radiometer (PPR) observations of Callisto. By fitting a single-$\Gamma$ model to spatially binned PPR observations, we obtain maps of thermal inertia and emissivity across the surface that reveal thermal inertiae between 15 and 500 increasing eastward from $\sim$105$^\circ$W toward $\sim$30$^\circ$W and anomalously high spectral emissivities co-located with Valhalla. Despite sparse coverage, we also find the region between $\sim$90$^\circ$W and $\sim$195$^\circ$W to have relatively high emissivities ($>$0.95). These thermal inertia maps complement the ALMA data, which are able only to return globally averaged thermal inertiae and emissivities without considerably greater temporal coverage and higher spatial resolution of each point across Callisto's surface. Additionally, extensive temporal coverage by the PPR data at Valhalla, Asgard, and the cratered plains revealed roughly constant emissivities but elevated thermal inertiae of 100--200 for Valhalla and Asgard compared to $\sim$20 for the cratered plains, another complementary result to the ALMA data which were unable to distinguish between high thermal inertiae and low emissivities at the former two features.

The ALMA observations presented here are the first published of Callisto's surface at 600--700 GHz and complement the first thermal maps of Callisto obtained at lower ALMA frequencies~\citep{camarca_multifrequency_2025}. Moreover, the PPR maps included in this work represent the first thermal property maps of Callisto's surface at any wavelength, complementing the global thermal properties derived using ALMA data in this work and by~\citet{camarca_multifrequency_2025}. Observations at multiple frequencies allow us to place robust constraints on the thermal properties of Callisto across both its relatively featureless cratered plains and major surface features such as Valhalla and Asgard.

Future science observations of Callisto using ALMA will both considerably enhance the spatial resolution of the derived thermal images and reduce uncertainties on inferred brightness temperatures, thereby paving the way for upcoming spacecraft missions. In particular, more extensive spatial and spectral coverage by the \textit{JUpiter ICy moons Explorer (JUICE)} spacecraft will allow us to extend these thermal property maps into the third dimension, thereby constraining the depth dependence of Callisto's thermal properties. Upcoming ground-based observatory upgrades will similarly allow us to improve upon this work. For instance, near-future upgrades to ALMA’s wideband sensitivity will improve the precision of brightness temperature measurements and in the longer term, the planned next-generation VLA (ngVLA) will provide even greater sensitivity and spatial resolution, enabling much tighter constraints on the spatial variability of Callisto’s thermal properties.

\vspace{0.5cm}

We acknowledge support from the National Science Foundation through a Graduate Research Fellowships under Grants No. DGE-2137419 to C.M. and No. DGE‐1745301 to M.C., as well as through grant 2308280, which supported C.M., M.C., K.d.K., and A.E.T. This research was also funded in part by the Heising-Simons Foundation through grant \#2019-1611. Funding was additionally provided by the NASA ROSES Solar System Observations program (through Task Order 80NM0018F0612) for A.E.T. and K.d.K. This paper makes use of the following ALMA data: ADS/JAO.ALMA\#2011.0.00199.S, ADS/JAO.ALMA\#2011 .0.00647.S, and ADS/ JAO.ALMA\# 2011.0.00223.S. ALMA is a partnership of ESO (representing its member states), NSF (USA) and NINS (Japan), together with NRC (Canada), MOST and ASIAA (Taiwan), and KASI (Republic of Korea), in cooperation with the Republic of Chile. The Joint ALMA Observatory is operated by ESO, AUI/NRAO, and NAOJ. The National Radio Astronomy Observatory is a facility of the National Science Foundation operated under cooperative agreement by Associated Universities, Inc. We thank Cole Wampler and Dominic Ludovici on the ALMA staff for staging and consulting on the quality of our calibration data, respectively. Finally, 21 PDS data products from the Photopolarimeter-Radiometer (PPR) RDR Data Set were used for the analysis contained within this paper. We thank the \textit{Galileo} project and PPR team for acquiring those data.


\vspace{5mm}
\facilities{ALMA, PDS, Galileo (PPR)}


\software{CASA~\citep{the_casa_team_casa_2022},
          JPL Horizons Ephemeris~\citep{rhodes_jplephem_2019},
          astropy~\citep{robitaille_astropy_2013},
          scipy~\citep{virtanen_scipy_2020},
          funkyfresh~\citep{milby_funkyfresh_2022}}


\appendix

\section{Robustness Metric}\label{section:robustness}

The $\chi^2$ value corresponding to the diurnal curve fitted to each PPR data bin represents the ability for the curve to fit the available measurements but importantly, does not necessarily represent the ability for the model to constrain the thermal inertia and emissivity values. For example, in cases where only one PPR measurement is available, the diurnal curve could take on any thermal inertia value and set the emissivity such that the curve passes through the PPR measurement, thereby yielding a $\chi^2$ of zero. Cases with two or three points are similarly easier to fit to a wide range of thermal inertiae and emissivities than cases with more points. To effectively trace the robustness of each curve fitting, we define the ``robustness metric'' $\mathcal{R}$ as follows:
\begin{equation}
    \mathcal{R} = \left[\frac{\overline{N}}{N_\text{max}}\right]\left[1 - \frac{\overline{w}}{w_\text{max}}\right],
\end{equation}
where $\overline{N}$ is the map of the number of PPR observations and $N_\text{max}$ is the map-wide maximum number of PPR observations. The thermal inertia width $w$ is defined as the width of the range of thermal inertiae that satisfy $2\chi^2_\text{min}$ at each bin (e.g., $w=385$ for well fitting thermal inertia of 15 to 400). Correspondingly, $\overline{w}$ is the map of thermal inertia widths and $w_\text{max}$ is the map-wide thermal inertia width maximum. Greater numbers of PPR observations and smaller thermal inertia widths each lead to enhanced robustness; as such, we define the robustness metric $\mathcal{R}$ to range from 0 to 1, with values near 1 corresponding to highly robust fits.

\begin{figure}[t]
\centering
\includegraphics[width=0.49\textwidth]{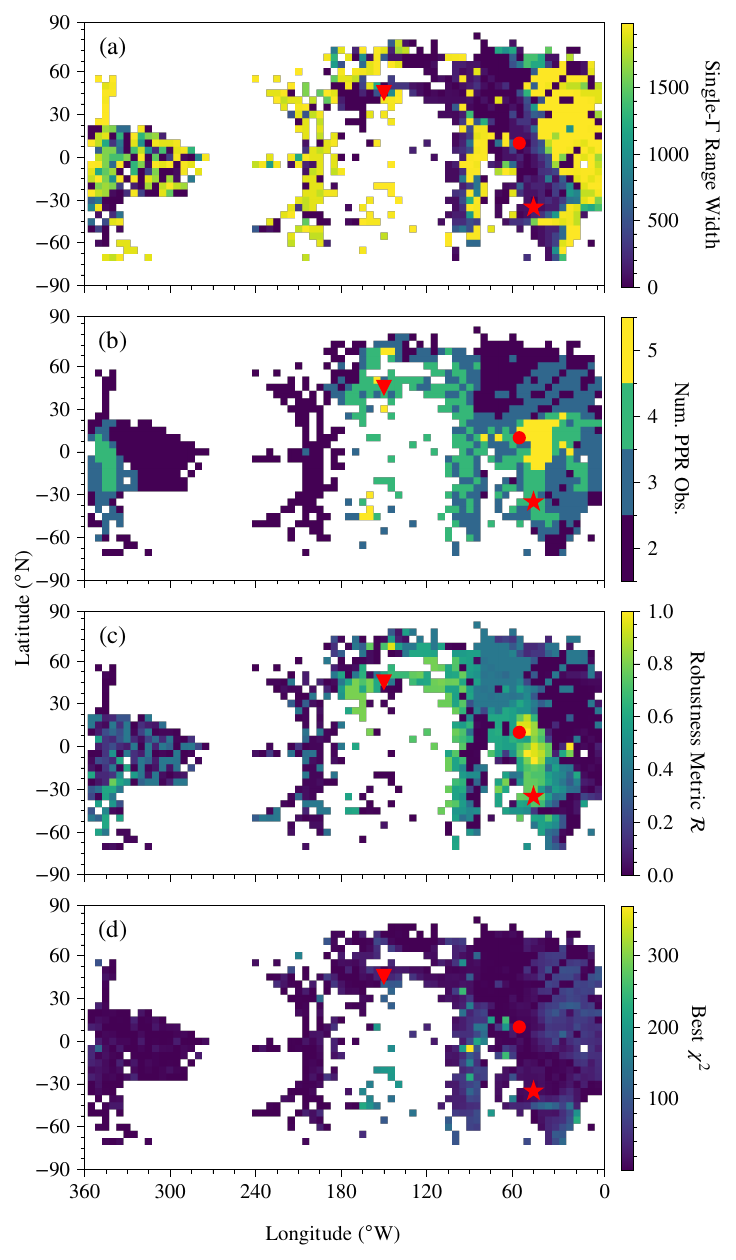}
\caption{Maps of metrics that may represent goodness of fit in different ways. (a) Width of the range of thermal inertiae that satisfy $\chi^2 < 2\chi^2_\text{min}$. (b) Number of PPR observations with coverage at each bin. (c) Robustness parameter $\mathcal{R}$. (d) $\chi^2_\text{min}$ at each bin. Each subplot uses its own color scale. The red circle, triangle, and star correspond to Valhalla, Asgard, and cratered plains, respectively. Unlike Figure~\ref{fig:PPRmaps}, this diagnostic figure includes pixels with robustness parameters $\mathcal{R}=0$.}
\label{fig:PPRappendix}
\end{figure}

\bibliography{refs}{}
\bibliographystyle{aasjournal}



\end{document}